\documentclass[aps,prd,onecolumn,groupedaddress,showpacs,nofootinbib,amssymb]{revtex4}
\usepackage[dvips]{graphicx}
\usepackage{amssymb}
\usepackage{amsmath}
\usepackage{graphicx}
\usepackage{amsfonts}
\usepackage{bm}

\begin{document}

\title{ Inflation in exponential scalar model and finite-time singularity induced instability}
\author{
S.~D.~Odintsov,$^{1,2,4}$\,\thanks{odintsov@ieec.uab.es}
V.~K.~Oikonomou,$^{3,4}$\,\thanks{v.k.oikonomou1979@gmail.com}}
\affiliation{ $^{1)}$Institut de Ciencies de lEspai (IEEC-CSIC),
Campus UAB, Carrer de Can Magrans, s/n\\
08193 Cerdanyola del Valles, Barcelona, Spain\\
$^{2)}$ ICREA, Passeig LluA­s Companys, 23,
08010 Barcelona, Spain\\
$^{3)}$ Department of Theoretical Physics, Aristotle University of
Thessaloniki,
54124 Thessaloniki, Greece\\
$^{4)}$ National Research Tomsk State University, 634050 Tomsk, Russia and Tomsk State Pedagogical University, 634061 Tomsk, Russia\\
}

\begin{abstract}

We investigate how a Type IV future singularity can be included in
the cosmological evolution of a well-known exponential model of
inflation. In order to achieve this we use a two scalar field model,
in the context of which the incorporation of the Type IV singularity
can be consistently done. In the context of the exponential model we study, when a Type IV singularity is included in the evolution, an instability occurs in the slow-roll parameters, and in particular on the second slow-roll parameter. Particularly, if we abandon the
slow-roll condition for both the scalars we shall use, then the most
consistent description of the dynamics of the inflationary era is
provided by the Hubble slow-roll parameters $\epsilon_H$ and
$\eta_H$. Then, the second Hubble slow-roll parameter $\eta_H$,
which measures the duration of the inflationary era, becomes
singular at the point where the Type IV singularity is chosen to
occur, while the Hubble slow-roll parameter $\epsilon_H$ is regular
there. Therefore, this infinite singularity indicates that the
occurrence of the finite-time singularity is responsible for the instability in the scalar field model we study. This sort
of instability has it's imprint on the dynamical system that can be
constructed from the cosmological equations, with the dynamical
system being unstable. Also the late-time evolution of the two
scalar field system is studied, and in the context of the
theoretical framework we use, late-time and early-time acceleration
are described in a unified way. In addition, the instability due to the singularity mechanism we propose, is discussed in the
context of other inflationary scalar potentials. Finally, we discuss the implications of such a singularity in the Hubble slow-roll parameters and we also critically discuss qualitatively, what implications could this effect have on the graceful exit problem of the exponential model.
\end{abstract}

\pacs{04.50.Kd, 95.36.+x, 98.80.-k, 98.80.Cq,11.25.-w}

\maketitle



\def\pp{{\, \mid \hskip -1.5mm =}}
\def\cL{\mathcal{L}}
\def\be{\begin{equation}}
\def\ee{\end{equation}}
\def\bea{\begin{eqnarray}}
\def\eea{\end{eqnarray}}
\def\tr{\mathrm{tr}\, }
\def\nn{\nonumber \\}
\def\e{\mathrm{e}}

\section{Introduction}

One of the unavoidable features of many cosmological theoretical
models is the appearance of cosmological singularities, which either
signal the need for an alternative physical description or maybe can
be viewed as the link between a classical and a quantum cosmological
theory. In cosmology singularities are considered according to two
alternative classification schemes, the one that uses the scale
factor as a rule for the classification
\cite{barrowsing1,barrowsing2,barrow,Barrow:2015ora} and another
classification scheme that uses the scale factor, the effective
energy density, the effective pressure and in some cases the Hubble
rate in order to determine the type of the singularity
\cite{Nojiri:2005sx,ref5}. The most interesting singularities which
are typical for quintessence/phantom dark energy evolution are
finite time future cosmological singularities and they were firstly
developed in \cite{barrowsing1} and in \cite{Nojiri:2005sx,sergnoj}
respectively. The most severe, and rather unwanted types of
singularities in every theoretical physics framework, are the
crushing type singularities, like the initial singularity or for
example the Big Rip \cite{Nojiri:2005sx,ref5}. For these
singularities, the Hawking-Penrose strong energy theorems
\cite{hawkingpenrose} are violated and therefore geodesics
incompleteness occurs. These singularities are possibly indicators
of the inability of the theoretical framework that predicts them, to
fully describe the physical evolution consistently at the spacetime
points that these occur. So probably these can be viewed as possible
indicators of the compelling need of a more fundamental theory for
the consistent description of the physical phenomena near these
spacetime points. For some informative studies on this kind of
singularities, see
\cite{Virbhadra:1995iy,Virbhadra:1998kd,Virbhadra:2002ju} and
references therein. Also for insightful studies on the avoidance of
the initial singularity see \cite{ellistsagas,bounce}. There exist
however other types of singularities that can be characterized as
''mild'' singularities, with mild indicating the fact that no
geodesics incompleteness occurs at these spacetime singularities and
also the Hawking-Penrose theorems, and even the weak energy theorems
are not violated. One characteristic example of this type of
singularities is the Type IV singularity, which can be viewed as a
link between the classical and the quantum description (whatever it
may be) of the theory under study.  Some properties and also
phenomenological-observational implications of Type IV singularities
as applied mainly to inflation, were recently studied in
\cite{noo1,noo2,noo3,noo4}.

Particularly, in Ref. \cite{noo3}, using standard reconstruction
schemes for scalar-tensor theories
\cite{Nojiri:2005pu,Capozziello:2005tf,Ito:2011ae}, we incorporated
a Type IV singularity in a model of $R^2$ inflation, by using two
scalar fields. As we demonstrated, the second scalar field has no
effect in the early-time dynamical evolution of the physical system
and contributes only at late-times. The purpose of this paper is two
fold: Firstly we shall demonstrate how we can consistently
incorporate a Type IV singularity in a quite well known exponential
scalar model of inflation
\cite{barrowexp,wettericg,encyclopedia,exponentialmodelofinflation,sahni}
of the form $V(\varphi )=V_0e^{-\gamma \varphi}$, by using two
scalar fields. Secondly, we shall demonstrate that a singularity occurs in the Hubble slow-roll parameters of the exponential scalar model with potential $V(\varphi
)=V_0e^{-\gamma \varphi}$. Note that according to the Planck data
\cite{planck} the model we shall study is consistent to some extent
to the measured observational data, but there is no exit from
inflation for this model, so the instability we found can maybe have an impact on the graceful exit problem of the exponential model.

Using two scalar fields, we shall
demonstrate that the theory has a source of instability for this scalar model. Particularly,
the main reason behind this instability is the Type IV singularity. As
we shall evince, owing to the Type IV singularity, the theory
becomes unstable because the slow-roll parameters and specifically
the second slow-roll parameter develops an infinite singularity at
the Type IV singularity point. Then, by suitably choosing the time
at which the Type IV singularity occurs, we can render the system unstable at any point we might wish. Note that the second
slow-roll index measures how long the inflationary era is, while the
first slow-roll parameter indicates if inflation occurs. In the
context of the two scalar field model we will work on, the second
scalar has a negligible effect at the early time evolution, apart
from the instability that it causes via the Type IV singularity, and
the early time evolution is solely governed by the canonical scalar
field with scalar potential the exponential potential of the form
$V(\varphi )=V_0e^{-\gamma \varphi}$. In addition, the second scalar
will eventually control the late-time era, although this can be
strongly model dependent. One of the important assumptions we shall
make is that the canonical scalar field corresponding to the
exponential potential $V(\varphi )=V_0e^{-\gamma \varphi}$, does not
satisfy the slow-roll conditions, in contrast to the case studied in
our previous study \cite{noo3}, in which the slow-roll conditions
were determined by the potential. Therefore, in that case, the
singularity did not modify or affect the slow-roll conditions,
unless the time that the singularity occurs was chosen before the
inflation ending time. The definition of the slow-roll parameter
using the potential was described in detail in \cite{barrowslowroll}
and was firstly given in \cite{liddleslowroll}. In contrast, the
slow-roll condition can involve the Hubble rate, as was firstly done
in \cite{copelandslowroll}, and we adopt this approach in this
paper. As we shall demonstrate, the singularity in the second
slow-roll index is due to the appearance of the second derivative of
the Hubble rate in it's functional form. With regards to the
dynamical system instability caused by a singularity there, we have
to note that this sort of instability is somehow different from the
infinite instabilities occurring in the effective equation of state
of single scalar field models when the phantom divide is crossed.
These instabilities were firstly pointed out in \cite{vikman} and
were studied in detail in \cite{Nojiri:2005pu,ref2}. Finally, we
demonstrate that in the context of the two scalars model we shall
use, it is possible to unify the early-time with late-time
acceleration.

This paper is organized as follows: In section II we present in
brief the classification of finite time singularities and also the
geometric background conventions we shall use throughout the rest of
the paper. Moreover, we describe the basic features of the
exponential scalar model, including the observational predictions
that it implies. In addition, we present the two scalar field model
and study it's dynamical evolution by appropriately constraining the
second scalar field. In section III we present in detail the two
sources of instability and also we qualitatively discuss the potential implications that the instability could have on the graceful exit problem of the exponential potential. Finally, the late-time dynamics of the two scalars model
is studied in section IV and in section V we present the implications of the Type IV singularities to other scalar models. The concluding remarks along with a
discussion of the results follow at the end of the paper.

\section{Exponential two scalar model of inflation and Type IV singular evolution}

As we already mentioned, the detailed classification of finite time
singularities was firstly developed in
Refs.~\cite{Nojiri:2005sx,sergnoj}, and we now present in brief the
essential features of this classification scheme. \begin{itemize}
\item Type I (``The Big Rip Singularity''): This singularity occurs when
the cosmic time approaches $t \to t_s$, where the following physical
quantities behave as: the scale factor $a$, the effective energy
density $\rho_{\mathrm{eff}}$ and finally the effective pressure
$p_\mathrm{eff}$ diverge, that is, $a \to \infty$,
$\rho_\mathrm{eff} \to \infty$, and $\left|p_\mathrm{eff}\right| \to
\infty$. This singularity is of crushing type, and more details can
be found in Refs.~\cite{Nojiri:2005sx,ref5}
\item Type II (``The Sudden Singularity''): This singularity occurs
when the cosmic time approaches $t \to t_s$, where only the scale
factor $a$ and the effective energy density $\rho_{\mathrm{eff}}$
tend to a finite value, that is, $a \to a_s$, $\rho_{\mathrm{eff}}
\to \rho_s$, but the effective pressure diverges,
$\left|p_\mathrm{eff}\right| \to \infty$. For details on this
singularity, consult Refs. \cite{barrowsing2,barrow}.
\item Type III : This singularity occurs when the cosmic time approaches $t \to t_s$,
where only the scale factor tends to a finite value, that is, $a \to
a_s$, but the rest two physical quantities, that is, the effective
pressure and the effective energy density, both diverge, that is,
$\left|p_\mathrm{eff}\right| \to \infty$ and $\rho_\mathrm{eff} \to
\infty$.
\item Type IV : This singularity is not of the crushing type and it is the most ''mild'' among
the four types of finite time cosmological singularities. In this
case, as $t \to t_s$, all the aforementioned quantities tend to
finite values, that is, $a \to a_s$, $\rho_\mathrm{eff} \to \rho_s$,
$\left|p_\mathrm{eff}\right| \to p_s$, and also the Hubble rate and
it's first derivative are finite. But the second or higher
derivatives of the Hubble rate diverge as $t \to t_s$. For details
on this type of singularity see Ref. \cite{Nojiri:2005sx}.
\end{itemize}

Having presented the essentials of the finite time cosmological
singularities classification, it is worth presenting also in brief
the geometrical conventions we shall use in the rest of the paper.
We shall assume that the background spacetime metric is a flat
Friedmann-Robertson-Walker (FRW), with line element of the form, \be
\label{JGRG14} ds^2 = - dt^2 + a(t)^2 \sum_{i=1,2,3}
\left(dx^i\right)^2\, , \ee where the parameter $a(t)$ stands for
the scale factor. Also, the effective energy density
$\rho_\mathrm{eff} $ and the effective pressure of the matter fluids
$p_\mathrm{eff}$ for the FRW of the form (\ref{JGRG14}) are related
to the Hubble rate in the following way, \be \label{IV}
\rho_\mathrm{eff} \equiv \frac{3}{\kappa^2} H^2 \, , \quad
p_\mathrm{eff} \equiv - \frac{1}{\kappa^2} \left( 2\dot H + 3 H^2
\right)\, . \ee

Originally, the finite-time future singularities were studied after
dark energy era. However, recently it was realized that they are
relevant and some of them maybe quite realistic also during or just
after inflation. In our previous studies \cite{noo1,noo2,noo3,noo4}
we incorporated a Type IV singularity in the cosmological (mainly
inflationary) evolution of some scalar-tensor models using one
scalar \cite{noo1} or two scalars \cite{noo2,noo3}. But in the case
of the nearly $R^2$ inflation potential which we studied in
\cite{noo3}, the use of two scalar fields was compelling, for
reasons we explained in detail in \cite{noo3}. The same reasoning
applies in the case we study in this paper too, therefore this
incorporation can be realized by using two scalar fields (details on
the reasoning why to use two scalar fields are presented in the
appendix). Before going into the details of our model, we shall
present in brief the exponential model of inflation
\cite{barrowexp,wettericg,encyclopedia,exponentialmodelofinflation,sahni}
and we also describe when this model can be partially compatible
with Planck data. Also we shall demonstrate why it is impossible to
have a graceful exit from the inflationary era, regardless if we
assume slow-roll or not.

The exponential model we shall be interested in, is a power-law
model of inflation \cite{encyclopedia}, in which, the action of the
canonical scalar field is the following, \be \label{ma7} S=\int d^4
x \sqrt{-g}\left\{ \frac{1}{2\kappa^2}R - \frac{1}{2}\partial_\mu
\varphi
\partial^\mu\varphi - V(\varphi)  \right\}\, ,
\ee with the scalar potential being of the following form,
\begin{equation}\label{scalarpotential}
V(\varphi )=V_0e^{-\sqrt{\frac{2}{q}}\kappa \varphi}\, ,
\end{equation}
and with $\kappa^2= 8\pi G$, where $G$ is Newton's constant. The
exponential potential of Eq. (\ref{scalarpotential}), has been used
in many cosmological contexts and for an incomplete list on this see
\cite{barrowexp,wettericg,encyclopedia,exponentialmodelofinflation,sahni}.
As we shall present in detail later on, there exist two alternative
descriptions for the slow-roll parameters, one that takes into
account the potential \cite{barrowslowroll,copelandslowroll}, and
the other one takes into account the Hubble rate only
\cite{barrowslowroll,liddleslowroll}. By taking into account only
the scalar potential, the slow-roll parameters are defined as
follows,
\cite{barrowslowroll,copelandslowroll,inflation,inflationreview},
\begin{equation}\label{slowrolpparaenedefs}
\epsilon =\frac{1}{2\kappa^2}\left(\frac{V '(\varphi )}{V (\varphi
)}\right),{\,}{\,}{\,}\eta
=\frac{1}{\kappa^2}\left(\frac{V''(\varphi )}{V (\varphi )}\right)
\, .
\end{equation}
The observational indices of inflation can be defined in terms of
these slow-roll parameters, and we shall be interested in the
spectral index of primordial curvature perturbations $n_s$ and the
tensor-to-scalar ration $r$. These are given in terms of the
slow-roll parameters in the following way
\cite{inflation,inflationreview},
\begin{equation}\label{observquantit}
n_s\sim 1-6\epsilon+2\eta\, , \quad r=16\epsilon\, , \quad a_s\sim
16 \epsilon \eta-24\epsilon^2-2\xi^2\, .
\end{equation}
For the potential (\ref{scalarpotential}), the slow-roll indices
read,
\begin{equation}\label{actpotslowroll}
\epsilon =\frac{1}{q},{\,}{\,}{\,}\eta =\frac{2}{q}\, ,
\end{equation}
which are called in the literature ''potential slow-roll
parameters'' \cite{barrowslowroll}, so we adopt this terminology
too. The corresponding observational indices are easily calculated,
and these read,
\begin{equation}\label{obindactval}
n_s\simeq 1-\frac{2}{q},{\,}{\,}{\,}r=\frac{16}{q}\, .
\end{equation}
However, in this paper we shall not assume that the slow-roll
approximation holds true, and therefore we need an alternative
description for the corresponding slow-roll parameters that measure
the inflationary dynamics. As was discussed in
\cite{barrowslowroll}, the slow-roll parameters of Eq.
(\ref{actpotslowroll}), when these are small, provide a necessary
consistency condition for inflation to occur, but not a sufficient
one. A much more superior definition of the slow-roll parameters
\cite{barrowslowroll} that encompass all the inflationary dynamics,
are the Hubble slow-roll parameters $\epsilon_H$ and $\eta_H$
\cite{barrowslowroll,copelandslowroll}, which are defined in terms
of the Hubble parameter as follows,
\cite{barrowslowroll,copelandslowroll},
\begin{equation}\label{hubbleslowrollparameters}
\epsilon_H =\frac{2}{\kappa^2}\left(\frac{H '(\varphi )}{H (\varphi
)}\right)^2,{\,}{\,}{\,}\eta_H =\frac{2}{\kappa^2}\frac{H''(\varphi
)}{H (\varphi )} \, ,
\end{equation}
where again we used the terminology of \cite{barrowslowroll}. It is
more convenient for the purposes of our analysis to express the
Hubble slow-roll parameters in terms of time and following
\cite{barrowslowroll}, these are given by,
\begin{equation}\label{slowrollprecond}
\epsilon_H
=3\frac{\frac{\dot{\varphi}}{2}}{V(\varphi)+\frac{\dot{\varphi}}{2}},{\,}{\,}{\,}\eta_H
=-3\frac{\ddot{\varphi}}{3 H\dot{\varphi}} \, .
\end{equation}
Using the following equations of motion for the canonical scalar
field $\varphi$,
\begin{equation}\label{canonicaleqnsmotion}
H^2=\frac{8\pi \kappa^2}{3}\left(\frac{\dot{\varphi}^2}{2}+V(\varphi
)\right),{\,}{\,}{\,}\ddot{\varphi}+3H\dot{\varphi}+V'(\varphi)=0\,
,
\end{equation}
the Hubble slow-roll parameters (\ref{slowrollprecond}) can be
written in the following way,
\begin{equation}\label{hubbleslowrolloriginal}
\epsilon_H =-\frac{\dot{H}}{H^2},{\,}{\,}{\,}\eta_H
=-\frac{\ddot{H}}{2H\dot{H}} \, ,
\end{equation}
In the rest of the paper we will make use of the Hubble slow-roll
parameters defined in Eq. (\ref{hubbleslowrolloriginal}), which are
suitable for our analysis. Note that in our previous work, related
to the $R^2$ inflation potential, we made use of the potential
slow-roll parameters, but in that case, the graceful exit was
guaranteed by the breaking of the slow-roll conditions, a very well
known mechanism \cite{linde}. For the potential
(\ref{scalarpotential}), the Hubble slow-roll indices read,
\begin{equation}\label{spedicslowrollind}
\epsilon = \frac{1}{q},{\,}{\,}{\,} \eta =\frac{1}{q},
\end{equation}
and the spectral index of primordial curvature perturbations reads
\cite{sahni},
\begin{equation}\label{sahnins}
n_s=1-\frac{1}{q-1}\, .
\end{equation}
The latest Planck data indicate that the values of the
aforementioned spectral indices are constrained as follows,
\begin{equation}\label{constraintedvalues}
n_s=0.9644\pm 0.0049\, , \quad r<0.10\, ,
\end{equation}
so the scalar-to-tensor ratio is already excluded (see also
\cite{sahni}), but the index $n_s$ can be compatible to the 2015
Planck data \cite{planck}, if the parameter $q$ takes the following
values,
\begin{equation}\label{interv}
0\lesssim q \lesssim 101\, ,
\end{equation}
so we choose $q=100$ for our analysis. Before we proceed, let us
discuss the issue of graceful exit from inflation for the model
(\ref{scalarpotential}). By looking at the potential slow-roll
parameters and also at the Hubble slow-roll parameters, in both
cases it is not possible to have graceful exit since these never
become of order one, and also in the context of a single scalar
field model, there is no alternative mechanism to trigger the
graceful exit from inflation. We have to mention that in the
literature there are two very popular mechanisms that can trigger
the exit from inflation, with the first being related to tachyonic
instabilities \cite{encyclopedia} and in the context of the second
mechanism, the graceful exit is triggered by a trace anomaly
\cite{sergeitraceanomaly}.

In this paper, we shall present a mechanism that renders the system dynamically unstable, with the instability being triggered by the Type IV
singularity. This mechanism requires the use of two scalar fields,
as we now demonstrate. In the appendix we provide the details on why
it is very difficult to incorporate the Type IV singularity to the
single scalar field exponential model with potential
(\ref{scalarpotential}).

In order to incorporate the Type IV singularity in the cosmological
model described partially by a canonical scalar field with scalar
potential that of Eq. (\ref{scalarpotential}), we shall use a two
scalar field scalar-tensor model. In principle, in the context of
two or more scalar field models, there are much more options for a
successful description-realization of various cosmologies. In
addition, the two scalar field models can remedy certain
inconsistencies that arise in the single scalar field models, mainly
having to do with the effective equation of state (EoS) of the
scalar field when the phantom divide crossing occurs
\cite{Nojiri:2005pu,Capozziello:2005tf,vikman}. Among the vast
possibilities we can choose for the double scalar field model, we
shall make a convenient choice that proves to be valuable from a
phenomenological point of view.

Consider the following two scalar field action, \be \label{A1}
S=\int d^4 x \sqrt{-g}\left\{\frac{1}{2\kappa^2}R
 - \frac{1}{2}\omega(\phi)\partial_\mu \phi \partial^\mu \phi
 - \frac{1}{2}\eta(\chi)\partial_\mu \chi\partial^\mu \chi
 - V(\phi,\chi)\right\}\, .
\ee where as usual, $\omega(\phi)$ stands for the kinetic function
of the non-canonical scalar field $\phi$, while $\eta (\chi)$
denotes the kinetic function of the second non-canonical scalar
field $\chi$. Of course it is conceivable that when one of the
functions $\omega(\phi)$ or  $\eta (\chi)$ is negative, then the
associated to this function scalar field becomes a phantom field,
but as we shall see, by appropriately choosing the initial
conditions of the scalar field $\chi$, the possibility of a phantom
scalar field and consequently of partially phantom inflation
\cite{Liu:2012iba} is avoided. Notice however that this issue is
strongly parameter and model dependent. Assuming that the scalar
fields depend only on the cosmic time $t$, then for the FRW
cosmological background (\ref{JGRG14}), the FRW equations for the
action (\ref{A1}) read, \be \label{A2} \omega(\phi) {\dot \phi}^2 +
\eta(\chi) {\dot \chi}^2 = - \frac{2}{\kappa^2}\dot H\, ,\quad
V(\phi,\chi)=\frac{1}{\kappa^2}\left(3H^2 + \dot H\right)\, . \ee By
choosing the generalized scalar potential $V(\phi,\chi)$ and the
functions $\omega(\phi)$, $\eta(\chi)$ to satisfy the following
relations, \be \label{A3} \omega(t) + \eta(t)=-
\frac{2}{\kappa^2}f'(t)\, ,\quad
V(t,t)=\frac{1}{\kappa^2}\left(3f(t)^2 + f'(t)\right)\, , \ee we can
find an explicit and quite general solution of Eqs.~(\ref{A2}),
which is of the form, \be \label{A4} \phi=\chi=t\, ,\quad H=f(t)\, .
\ee Practically, what we just described is the scalar reconstruction
method, and for further details and applications, the reader is
referred to \cite{Nojiri:2005pu}. Since this method provides us with
much freedom for choosing the field kinetic functions, we choose
these as follows,

\be \label{A5} \omega(\phi) =-\frac{2}{\kappa^2}\left\{f'(\phi) -
\sqrt{\alpha(\phi)^2 + f'(\phi)^2} \right\}\, ,\quad \eta(\chi) =
-\frac{2}{\kappa^2}\sqrt{\alpha(\chi)^2 + f'(\chi)^2}\, . \ee where
the function $\alpha (x)$ is arbitrarily chosen and it's explicit
form will be specified shortly. The scalar potential can be
specified in terms of an auxiliary function $\tilde f(\phi,\chi)$,
which is defined to be equal to, \be \label{A6} \tilde
f(\phi,\chi)\equiv - \frac{\kappa^2}{2}\left(\int d\phi \omega(\phi)
+ \int d\chi \eta(\chi)\right)\, . \ee and is chosen to satisfy the
following, \be \label{A7} \tilde f(t,t)=f(t)\, . \ee
 Then, the scalar potential can be expressed in terms of the auxiliary function $\tilde f(\phi,\chi)$, as follows,
\be \label{A8} V(\phi,\chi)=\frac{1}{\kappa^2}\left(3{\tilde
f(\phi,\chi)}^2 + \frac{\partial \tilde f(\phi,\chi)}{\partial \phi}
+ \frac{\partial \tilde f(\phi,\chi)}{\partial \chi} \right)\, . \ee
Notice that, the constraint (\ref{A7}) fixes the values of the
integration constants that arise in Eq.~(\ref{A6}). The two
non-canonical scalar fields $\phi$ and $\chi$ satisfy the following
equations of motion, \be \label{A9} 0 = \omega(\phi)\ddot\phi +
\frac{1}{2}\omega'(\phi) {\dot \phi}^2 + 3H\omega(\phi)\dot\phi +
\frac{\partial \tilde V(\phi,\chi)}{\partial \phi} \, , \quad 0 =
\eta(\chi)\ddot\chi + \frac{1}{2}\eta'(\chi) {\dot \chi}^2 +
3H\eta(\chi)\dot\chi + \frac{\partial \tilde V(\phi,\chi)}{\partial
\chi}\, . \ee In order to consistently incorporate a finite time
cosmological singularity in the two scalar field cosmological
evolution, the Hubble parameter is assumed to have the following
general form, \be \label{IV1Bnoj1} H(t) = f_1(t) + f_2(t) \left( t_s
- t \right)^\alpha\, , \ee where the parameter $\alpha$ actually
determines the type of the finite time cosmological singularity, and
it will be assumed to take non-integer values of the form, \be
\label{IV2} \alpha= \frac{n}{2m + 1}\, , \ee with $m$ and $n$
integer numbers, the values of which will be determined shortly.
According to the classification of finite time cosmological
singularities we presented previously, the values of $\alpha $
classify the cosmological singularities of the evolution model
(\ref{IV1Bnoj1}) in the following way,
\begin{itemize}\label{lista}
\item $\alpha<-1$ corresponds to the Type I singularity.
\item $-1<\alpha<0$ corresponds to Type III singularity.
\item $0<\alpha<1$ corresponds to Type II singularity.
\item $\alpha>1$ corresponds to Type IV singularity.
\end{itemize}
Since we are interested in the Type IV singularity, for the rest of
this paper it is assumed that $\alpha>1$, so the integers $m$ and
$n$ appearing in (\ref{IV2}), can be chosen in such a way so that
$\alpha>1$. For the purposes of this paper, $\alpha$ is assumed to
be equal to $\alpha=4/3$, for reasons that will become obvious soon.

A convenient choice for the function $\alpha (x)$ which appears in
Eq. (\ref{A5}), is the following,
\begin{equation}\label{alphaxnewdef1}
\alpha (x)=\sqrt{\left ( f_2' (x)\left (t_s-x \right
)^{\alpha}+\alpha f_2(x)\left (t_s-x\right )^{\alpha-1}\right
)^2-\left( H'(x)\right)^2}\, .
\end{equation}
and it is conceivable that the variable $x$ stands for the scalar
fields $\phi$ or $\chi$. By using this form of $\alpha (x)$ and
substituting it to Eq. (\ref{A5}), the kinetic functions of the
non-canonical scalar fields become significantly simplified and are
equal to, \be \label{B1noj} \omega(\phi)
=-\frac{2}{\kappa^2}f_1'(\phi)\, ,\quad \eta(\chi) =
-\frac{2}{\kappa^2} \left( f_2'(\chi) \left( t_s - \chi
\right)^\alpha + \alpha f_2(\chi) \left( t_s - \chi \right)^{\alpha
- 1} \right)\, , \ee In addition, by using (\ref{alphaxnewdef1}),
the auxiliary function $\tilde{f}(\phi,\chi)$ of Eq. (\ref{A6}),
becomes, \be \label{B2noj} \tilde f(\phi,\chi) = f_1(\phi) +
f_2(\chi) \left( t_s - \chi \right)^\alpha\, , \ee and consequently,
the two scalar field potential $V(\phi,\chi)$ becomes, \be
\label{B3noj} V(\phi,\chi)=\frac{1}{\kappa^2}\left(3\left( f_1(\phi)
+ f_2(\chi) \left( t_s - \chi \right)^\alpha \right)^2 + f_1'(\phi)
+  f_2'(\chi) \left( t_s - \chi \right)^\alpha + \alpha f_2(\chi)
\left( t_s - \chi \right)^{\alpha - 1} \right)\, . \ee Equipped with
Eqs. (\ref{B1noj}) and (\ref{B3noj}), we can easily realize a
singular evolution related to the scalar potential
(\ref{scalarpotential}). Specifically, we choose the Hubble rate
(\ref{IV1Bnoj1}) to be equal to, \be \label{IV1Bnoj} H(t) =
\frac{c_1}{c_2+c_3 t}+ c_4 \left( t_s - t \right)^\alpha\, , \ee
with the parameters $c_i$, $i=1,...,4$, being for the moment
arbitrary constants to be specified soon. Using Eq. (\ref{IV1Bnoj}),
the function $\alpha (x)$ appearing in Eq. (\ref{alphaxnewdef1})
becomes equal to,
\begin{equation}\label{alphaxnewdef}
\alpha (x)=\sqrt{c_4^2 (t_s-x)^{-2+2 \alpha } \alpha
^2-\left(-\frac{c_1 c_3}{(c_2+c_3 x)^2}-c_4 (t_s-x)^{-1+\alpha }
\alpha \right)^2}\, .
\end{equation}
and also the kinetic functions $\omega (\phi)$ and $\eta (\chi)$ of
Eq. (\ref{B1noj}) become equal to,
\begin{equation}\label{omeganforhilltopdouble}
\omega (\phi )=\frac{2 c_1 c_3}{\kappa ^2 (c_2+c_3 \phi )^2}\, ,
\quad \eta (\chi)= -\frac{2 c_2 \alpha  (t_s-\chi )^{-1+\alpha
}}{\kappa ^2}\, .
\end{equation}
Notice that in principle, the scalar field $\chi$ can become a
phantom scalar, and indeed this can be the case \footnote{It is easy
to see this, since for $\alpha=4/3$, the function $\eta (\chi)$ is
equal to, $\eta (\chi)= -\frac{8 c_2  (t_s-\chi )^{1/3 }}{3\kappa
^2}=\frac{8 c_2  (\chi-t_s )^{1/3 }}{3\kappa ^2}$, which can be
phantom when $\chi<t_s$. For the values of the parameters that we
will choose however, this never occurs} when $\alpha=4/3$. However,
as we will demonstrate, the scalar field $\chi$ turns to phantom
much more earlier than the inflationary era. In addition, in view of
the choice (\ref{alphaxnewdef}), the auxiliary function $\tilde
f(\phi,\chi)$ takes the following form,
\begin{equation}\label{barf}
\tilde f(\phi,\chi)=\frac{c_1}{c_2+c_3 \phi }+c_4 (t_s-\chi
)^{\alpha }\, ,
\end{equation}
and therefore, the scalar potential of Eq. (\ref{B3noj}) reads,
\begin{align}\label{scalarpotentialhilltop}
& V(\phi,\chi)=\frac{3 c_1^2}{\kappa ^2 (c_2+c_3 \phi )^2}-\frac{c_1
c_3}{\kappa ^2 (c_2+c_3 \phi )^2}-\frac{c_4 \alpha  (t_s-\chi
)^{-1+\alpha }}{\kappa ^2}+\frac{6 c_1 c_4 (t_s-\chi )^{\alpha
}}{\kappa ^2 (c_2+c_3 \phi )}+\frac{3 c_4^2 (t_s-\chi )^{2 \alpha
}}{\kappa ^2}\, .
\end{align}
In order to obtain the scalar potential of Eq.
(\ref{scalarpotential}), we will transform the scalar field $\phi$
to a canonical scalar field $\varphi$, by making the transformation
\be \label{ma13} \varphi \equiv \int^\phi d\phi \sqrt{\omega(\phi)}
\,, \ee with $\omega (\phi)$ being of the form given in Eq.
(\ref{omeganforhilltopdouble}), which yields,
\begin{equation}\label{noncanonicalcanonicalrel}
c_2+c_3 \phi = e^{\kappa\sqrt{\frac{c_3}{2c_1}}\varphi }
\end{equation}
By using the canonical scalar field $\varphi$, the action of Eq.
(\ref{A1}) takes the following form, \be \label{A1new} S=\int d^4 x
\sqrt{-g}\left\{\frac{1}{2\kappa^2}R
 - \frac{1}{2}\partial_\mu \varphi \partial^\mu \varphi
 -\frac{1}{2}\left(\frac{2 c_2 \alpha  (\chi-t_s )^{-1+\alpha }}{\kappa ^2}\right)\partial_\mu \chi\partial^\mu \chi
 - \tilde{V}(\varphi,\chi)\right\}\, .
\ee with the scalar potential $\tilde{V}(\varphi,\chi)$ being equal
to,
\begin{align}\label{finalpotential}
& \tilde{V}(\varphi,\chi)= \left(\frac{3 c_1^2}{\kappa ^2
}-\frac{c_1 c_3 }{\kappa ^2 }\right)e^{-2\kappa\sqrt{\frac{c_3}{2c_1}}\varphi  } -\frac{c_4 \alpha  (t_s-\chi )^{-1+\alpha }}{\kappa
^2}+\frac{6 c_1 c_4 (t_s-\chi )^{\alpha } e^{-\kappa\sqrt{\frac{c_3}{2c_1}}\varphi }}{\kappa ^2 }+\frac{3 c_4^2 (t_s-\chi )^{2 \alpha
}}{\kappa ^2}\, ,
\end{align}
where, since $\alpha=4/3$, we used the following relation in Eq.
(\ref{A1new}),
\begin{equation}\label{neweqn}
(t_s-\chi )^{-1+\alpha }=-(\chi-t_s )^{-1+\alpha }
\end{equation}
It is obvious that the potential of Eq. (\ref{finalpotential})
contains the scalar potential of Eq. (\ref{scalarpotential}), plus
the potential of the second scalar field $\chi$. In order for the
exponential potential of Eq. (\ref{finalpotential}) to be identical
to the scalar potential of Eq. (\ref{scalarpotential}), we must
require that,
\begin{equation}\label{furtherconstriants}
V_0=\frac{3 c_1^2}{\kappa ^2 }-\frac{c_1 c_3 }{\kappa ^2
},{\,}{\,}{\,}\frac{c_1}{c_3}=q\, .
\end{equation}
At this point we shall make a critical assumption for the dynamical
evolution of the system of the two scalar fields $\varphi$ and
$\chi$, described by the potential (\ref{finalpotential}) and for a
Hubble rate given in (\ref{IV1Bnoj}). Particularly, we require the
following,

\begin{itemize}\label{listaitem}

\item The canonical scalar field $\varphi$ does not satisfy the slow-roll requirements.

\item The non-canonical scalar field $\chi$ does not satisfy the slow-roll requirements too.

\item The non-canonical scalar $\chi$ is assumed to have very small values at early times.

\item The parameters $c_i$, $i=1,..,4$ are chosen in such a way, so that at early times the
contribution of the scalar field $\chi$ to the potential
$\tilde{V}(\varphi,\chi)$ given
 in Eq. (\ref{finalpotential}) is negligible, and also the kinetic function of the scalar field $\chi$ is also negligible at early times.

\item The Type IV singularity, which occurs at the time $t_s$, is chosen to appear at an
early time $10^{-35}$sec, where it is usually assumed that inflation
ends \cite{tasi}.

\end{itemize}
By taking into account the requirements of the list above and
combine these with the constraints (\ref{furtherconstriants}), and
also that we assumed $q=100$, we choose the parameters $c_i$,
$i=1,..,4$ as follows,
\begin{equation}\label{paramvalues}
c_1=10^{12},{\,}{\,}{\,}c_2=10^{-20},{\,}{\,}{\,}c3=10^{10},{\,}{\,}{\,}c_4=10^{-40}
\end{equation}
Also we shall take $t_s$, to be equal to $t_s=10^{-35}$sec, which is
the time that inflation ends in most theories. Of course inflation
does not end in the present model for the moment, but as we describe
soon, there is a mechanism that can potentially create a dynamical instability and this instability could be an indicator of graceful exit.

\begin{figure}[h]
\centering
\includegraphics[width=20pc]{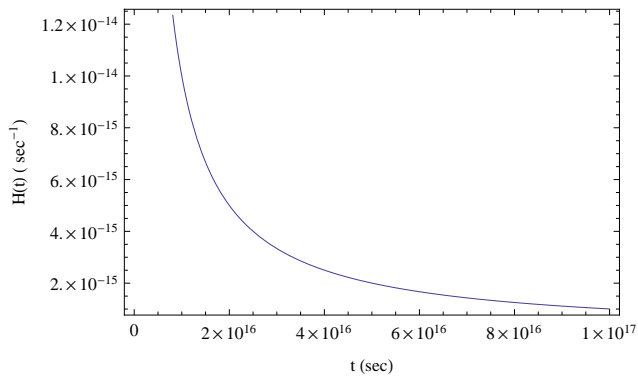}
\caption{\it{{The Hubble rate $H(t)=\frac{c_1}{c_2+c_3 t}+ c_4
\left( t_s - t \right)^\alpha$ as a function of the cosmic time with
$c_1=10^{12}$, $c_2=10^{-20}$, $c_3=10^{10}$, $c_4=10^{-40}$,
$t_s=10^{-35}$sec and $\alpha=4/3$.}}} \label{plot1}
\end{figure}

For the choice of the parameters (\ref{paramvalues}), the Hubble
rate is positive, as it can be seen in Fig. (\ref{plot1}).

Before continuing our analysis we need to discuss the issue of the physical units we shall adopt. For the purposes of our analysis, we shall adopt unit system for which $\hbar=c=1$, where $c$ is the speed of light. We shall choose to express every physical quantity in seconds. So the length $l$ is measured in seconds ($c=1$), time in seconds, mass is measured in sec$^{-1}$ (since $l=\hbar/mc=1/m$, when $c=1$). Therefore, the canonical scalar field is measured in sec$^{-1}$, the Hubble rate is measured in sec$^{-1}$, while the potential in sec$^{-4}$. Finally, the non-canonical scalar field is dimensionless, since $\kappa^2$ is measured in sec$^2$ and the derivative in sec$^{-1}$. In the following, we shall use these dimensions in our analysis and it is conceivable that all parameters can be expressed in these units.

For the values of the parameters (\ref{paramvalues}), and by
choosing a significantly small initial condition for the scalar
field $\chi$ at early times (earlier than $t_s=10^{-35}$sec) the
contribution of the scalar field to the potential
$\tilde{V}(\varphi,\chi)$ of Eq. (\ref{finalpotential}), is
negligible. Also the kinetic term of the scalar field $\chi$ is also
negligible at early times. We assume that the initial conditions for
the scalar field $\chi$ at $t=0$ is,
\begin{equation}\label{initialcondscalarchi}
\chi(0)=10^{-25},{\,}{\,}{\,}\chi'(0)=10\, ,
\end{equation}
for reasons which will become clear when we discuss about the
instability issue. Having all these considerations into mind, we
conclude that at early times, the scalar field action is governed by
the canonical scalar field $\varphi$, and is approximately equal to,
\be \label{approxma7} S=\int d^4 x \sqrt{-g}\left\{
\frac{1}{2\kappa^2}R - \frac{1}{2}\partial_\mu \varphi
\partial^\mu\varphi - V(\varphi)  \right\}\, ,
\ee
 with $V(\varphi)$ given in Eq. (\ref{scalarpotential}), since the term
  $+\frac{6 c_1 c_4 (t_s-\chi )^{\alpha } e^{-\kappa\sqrt{\frac{c_3}{2c_1}}\varphi }}{\kappa ^2 }$,
   and also the rest $\chi$-dependent terms in Eq. (\ref{finalpotential}), are negligible. In addition,
   as the cosmic time $t$ grows larger, the contribution of the second scalar field $\chi$ will eventually
   dominate the dynamics of the model, so that at late times, the scalar action is approximately equal to,
\be \label{A1newnew} S\simeq \int d^4 x
\sqrt{-g}\left\{\frac{1}{2\kappa^2}R
 - \frac{1}{2}\partial_\mu \varphi \partial^\mu \varphi
 +\frac{1}{2}\left(\frac{2 c_2 \alpha  (t_s-\chi )^{-1+\alpha }}{\kappa ^2}\right)\partial_\mu \chi\partial^\mu \chi
 - \tilde{V}_{\chi}(\varphi,\chi)\right\}\, .
\ee with $\tilde{V}_{\chi}(\varphi,\chi)$ being this time equal to,
\begin{equation}\label{newtildepot}
\tilde{V}_{\chi}(\varphi,\chi)\simeq -\frac{c_4 \alpha  (t_s-\chi
)^{-1+\alpha }}{\kappa ^2}+\frac{6 c_1 c_4 (t_s-\chi )^{\alpha }
e^{-\kappa\sqrt{\frac{c_3}{2c_1}}\varphi }}{\kappa ^2 }+\frac{3
c_4^2 (t_s-\chi )^{2 \alpha }}{\kappa ^2}\, .
\end{equation}
Depending on the dynamics and on which time we choose inflation to
end, it is possible that the $\varphi$-dependent exponential in
(\ref{newtildepot}) can be either equal to one, or it can be
approximately equal to zero. The qualitative picture in both cases
is the same, but the quantitative picture is different, with the
scalar field $\chi$ becoming dominant much more earlier in the case
that the $\varphi$-dependent potential term is included in the
analysis. So it is important to study both cases.

Before proceeding
to the two mechanisms that can generate the instability of the system under study, we study numerically the dynamical evolution of the
scalar field $\chi$, for the choice of parameters
(\ref{paramvalues}), in order to verify that our claims can be
realized in the model of the two scalars. We will show that the
scalar field $\chi$ provides a negligible contribution to the two
scalar fields model action at early times, but can dominate the
late-time evolution.

\begin{figure}[h]
\centering
\includegraphics[width=15pc]{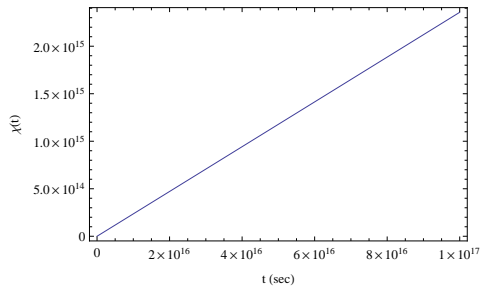}
\caption{\it{{The scalar field's $\chi (t)$ evolution as a function
of cosmic time $t$ with $c_1=10^{12}$, $c_2=10^{-20}$,
$c_3=10^{10}$, $c_4=10^{-40}$, $t_s=10^{-35}$sec, $\alpha=4/3$ and
for the initial conditions $\chi (0)\simeq 10^{-25}$, $\chi
'(0)\simeq 10$. }}} \label{plot3}
\end{figure}

\begin{figure}[h]
\centering
\includegraphics[width=15pc]{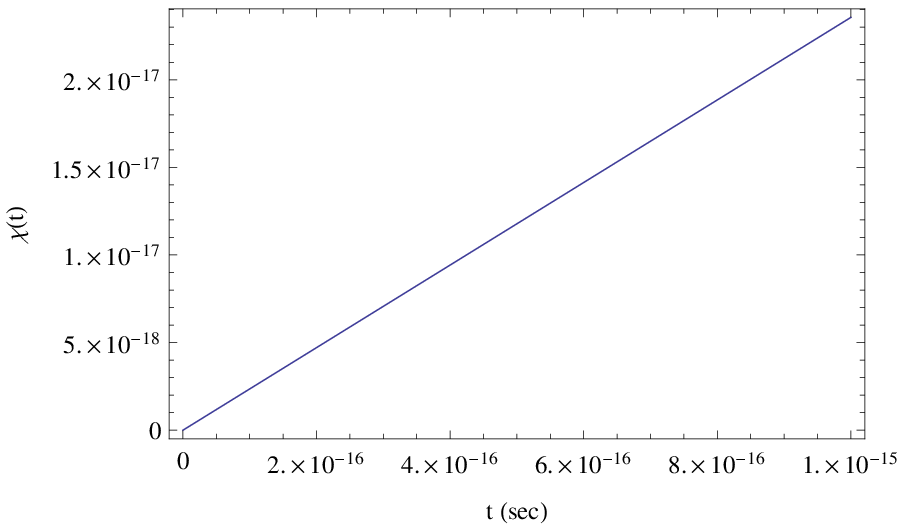}
\includegraphics[width=15pc]{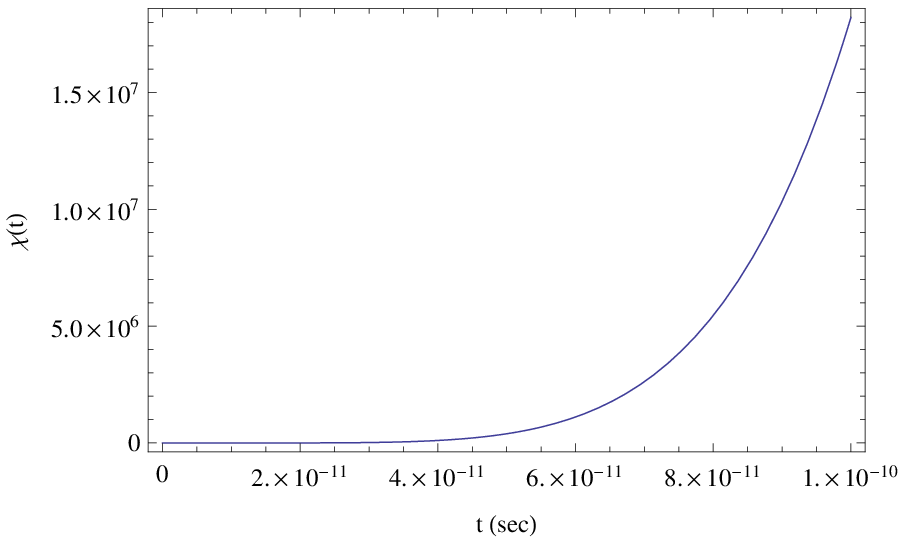}
\caption{\it{{The time dependence of the scalar field $\chi (t)$,
with $c_1=10^{12}$, $c_2=10^{-20}$, $c_3=10^{10}$, $c_4=10^{-40}$,
$t_s=10^{-35}$sec,$\alpha=4/3$ and for the initial conditions $\chi
(0)\simeq 10^{-25}$, $\chi '(0)\simeq 10$, including the
$\varphi$-dependent term. }}} \label{plot4}
\end{figure}

We intend to study in detail the dynamical evolution of the scalar
field $\chi$, by solving numerically the following equation of
motion,
\begin{equation}\label{nonslowrolleqnmotion}
\eta(\chi)\ddot\chi + \frac{1}{2}\eta'(\chi) {\dot \chi}^2 +
3H\eta(\chi)\dot\chi + \frac{\partial \tilde V(\phi,\chi)}{\partial
\chi}=0
\end{equation}
which governs the non slow-roll evolution of the scalar field
$\chi$. Recall that the potential $V(\varphi,\chi)$ is approximately
equal to,
\begin{equation}\label{approxmpot1}
V(\varphi,\chi)\simeq -\frac{c_4 \alpha  (t_s-\chi )^{-1+\alpha
}}{\kappa ^2}+\frac{6 c_1 c_4 (t_s-\chi )^{\alpha } e^{-\kappa\sqrt{\frac{c_3}{2c_1}}\varphi }}{\kappa ^2 }+\frac{3 c_4^2
(t_s-\chi )^{2 \alpha }}{\kappa ^2}\, ,
\end{equation}
if the scalar field $\varphi $ is assumed to have quite large values
at early times, so the term $+\frac{6 c_1 c_4 (t_s-\chi )^{\alpha }
e^{-\kappa\sqrt{\frac{c_3}{2c_1}}\varphi }}{\kappa ^2 }$ is
neglected. We assume that the scalar field $\chi$ satisfies the
initial condition (\ref{initialcondscalarchi}), in conjunction with
$\chi '(0)\simeq 10$, and that the parameters are chosen as in Eq.
(\ref{paramvalues}).
\begin{figure}[h]
\centering
\includegraphics[width=15pc]{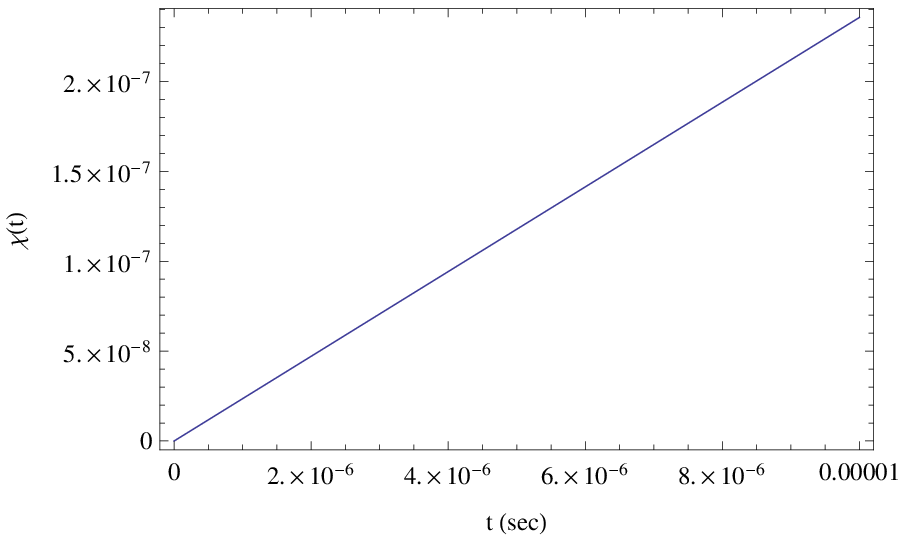}
\includegraphics[width=15pc]{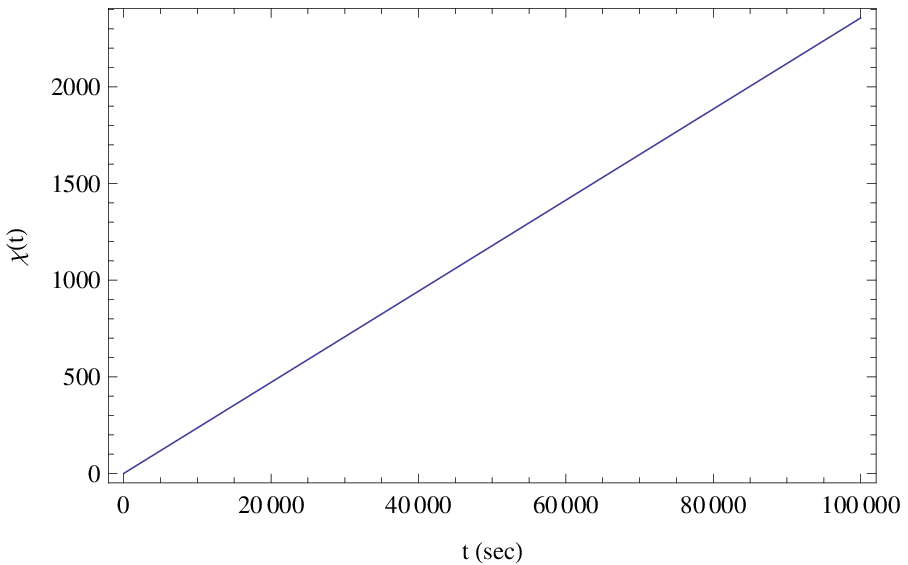}
\caption{\it{{The time dependence of the scalar field $\chi (t)$,
with $c_1=10^{12}$, $c_2=10^{-20}$, $c_3=10^{10}$, $c_4=10^{-40}$,
$t_s=10^{-35}$sec, $\alpha=4/3$ and for the initial conditions $\chi
(0)\simeq 10^{-25}$, $\chi '(0)\simeq 10$. }}} \label{plot2}
\end{figure}
 Also note that the present time in seconds is approximately $t_p\simeq 4.25\times 10^{17}$sec and
 also that $\kappa^2=8\pi G=2.0944\times 10^{-18}$GeV$^{-1}$, with $G$ denoting Newton's constant.
 Using these, in Figs. \ref{plot2} and \ref{plot3}, we present the time dependence of the scalar
 field $\chi$. As it is obvious, the scalar field $\chi$ is very small at early time, and remains
 small for a wide range of time, until present time where it grows large. After the present time era,
 it evolves larger and dominates the evolution. This qualitatively appealing behavior was also pointed
 out in \cite{noo3}, and we briefly address this issue in a later section.
It is worth checking whether the inclusion of the
$\varphi$-dependent potential term $\frac{6 c_1 c_4 (t_s-\chi
)^{\alpha } e^{-\kappa\sqrt{\frac{c_3}{2c_1}}\varphi }}{\kappa ^2
}$, in the scalar field $\chi$ potential results to a different
behavior. In the most extreme case, the $\varphi$-dependent term
will be of the form $\frac{6 c_1 c_4 (t_s-\chi )^{\alpha } }{\kappa
^2 }$, which occurs if $\varphi$ is approximately equal to zero
after inflation ends (assuming that our proposed mechanism ends
inflation eventually). However, even including this term, the
qualitative behavior remains the same, as it can be seen in Fig.
\ref{plot4}, but the scalar field $\chi$ dominates to a great extent
the evolution much more earlier in comparison to the other case.
The same analysis can be performed for the kinetic function $\eta
(\chi)$, which as we claimed previously, can be neglected at early
times. Indeed, in Table \ref{kinetictable}, we have calculated the
values of the kinetic function $\eta (\chi (t))$, for various time
moments and also for the cases that the $\varphi$-dependent
potential term is included (indicated by $\eta_w (\chi (t))$) or not
(indicated by $\eta_{n} (\chi (t))$).
\begin{table}[h]
\begin{center}
\begin{tabular}{|c|c|c|c|c|}
  \hline
  Time & $t\simeq 10^{-34}$ &  $t\simeq 10^{-15}$ & $t\simeq 10^{-5}$ & $t\simeq 10^{-4}$ \\
  \hline
  $\eta (\chi_n (t) )$ & $3.48\times 10^{-13}$ & $2.35\times 10^{-10}$ & $4.1 10^{-7}$ & $8.8\times 10^{-7}$ \\
  \hline
  $\eta (\chi_w (t) )$ & $3.47\times 10^{-13}$ & $2.2\times 10^{-10}$ & $8.8\times 10^{11}$ & $9.2\times 10^{13}$ \\
  \hline
\end{tabular}
\end{center}
\caption{\label{kinetictable}for the cases that the
$\varphi$-dependent potential term is included (indicated by $\eta_w
(\chi (t))$) or not (indicated by $\eta_{n} (\chi (t))$).}
\end{table}
As it can be seen from Table (\ref{kinetictable}), the behavior of
the kinetic function is quite similar at early times in both cases,
which are relevant for the inflationary era and in both cases the
kinetic function is negligible and consequently the same applies for
the kinetic term. The same analysis can be done for the potential of
the scalar field $\chi (t)$, both for the cases that the
$\varphi$-dependent term is kept in the potential or not. In Table
\ref{potentialchi}, we present the values of the scalar field $\chi$
potential, with the $\varphi$-dependent term included in the
potential (denoted as $\tilde{V}_{\chi}^w$) or not included (denoted
as $\tilde{V}_{\chi}^n$).
\begin{table}[h]
\begin{center}
\begin{tabular}{|c|c|c|c|c|}
  \hline
  Time & $t\simeq 10^{-34}$ &  $t\simeq 10^{-15}$ & $t\simeq 10^{-5}$ & $t\simeq 10^{-4}$ \\
  \hline
  $\tilde{V}_{\chi}^n$ & $1.5\times 10^{-13}$ & $1.6\times 10^{-10}$ & $3.5\times 10^{-7}$ & $7.73\times 10^{-7}$ \\
  \hline
  $\tilde{V}_{\chi}^w$ & $3.46\times 10^{-13}$ & $1.9\times 10^{-10}$ & $8.8\times 10^{11}$ & $9.2\times 10^{13}$ \\
  \hline
\end{tabular}
\end{center}
\caption{\label{potentialchi}The scalar field's $\chi (t)$ scalar
potential $\tilde{V}_{\chi}$ as a function of cosmic time $t$, with
the $\varphi$-dependent term included in the potential (denoted as
$\tilde{V}_{\chi}^w$) or not included (denoted as
$\tilde{V}_{\chi}^n$)}
\end{table}
As we can see in Table \ref{potentialchi}, the behavior of the
potential for early times, in case that the $\varphi$-dependent term
is included or not, is quite similar in both cases and also the
potential is negligible at early times.

\begin{figure}[h]
\centering
\includegraphics[width=15pc]{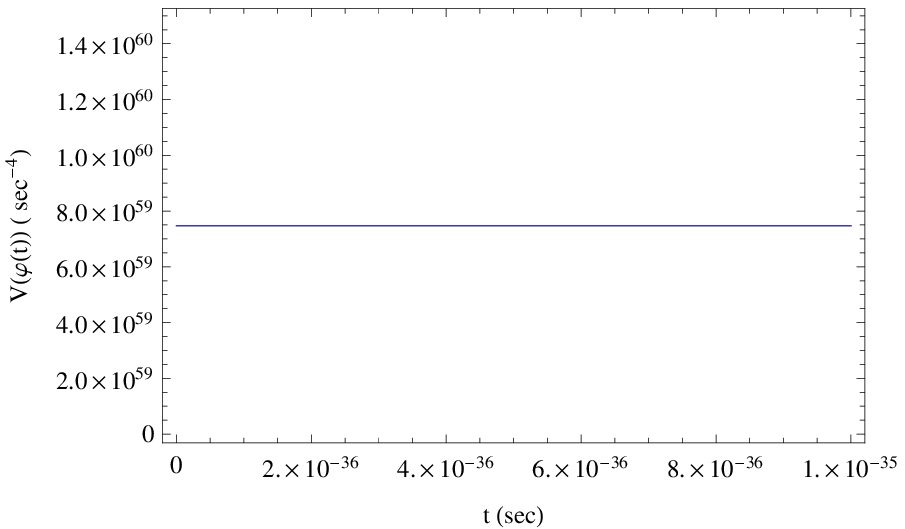}
\includegraphics[width=15pc]{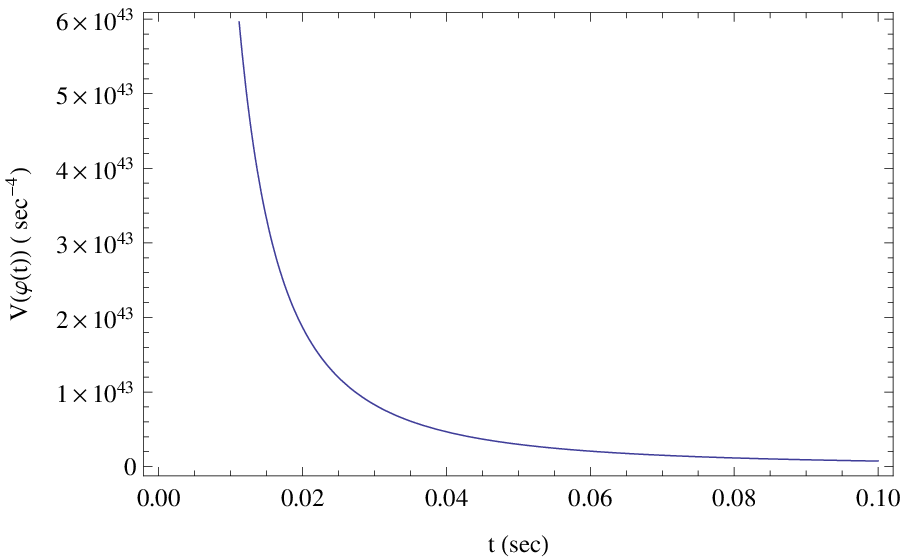}
\caption{\it{{The time dependence of the scalar field potential
$V(\varphi (t))$, with $c_1=10^{12}$, $c_2=10^{-20}$, $c_3=10^{10}$,
$c_4=10^{-40}$, $t_s=10^{-35}$sec,$\alpha=4/3$ and for the initial
conditions $\varphi (0)\simeq 10^{20}$ (sec$^{-1}$), $\varphi '(0)\simeq 10^{-2}$ (sec$^{-2}$),
including the $\varphi$-dependent term. }}} \label{plot5}
\end{figure}

\begin{figure}[h]
\centering
\includegraphics[width=15pc]{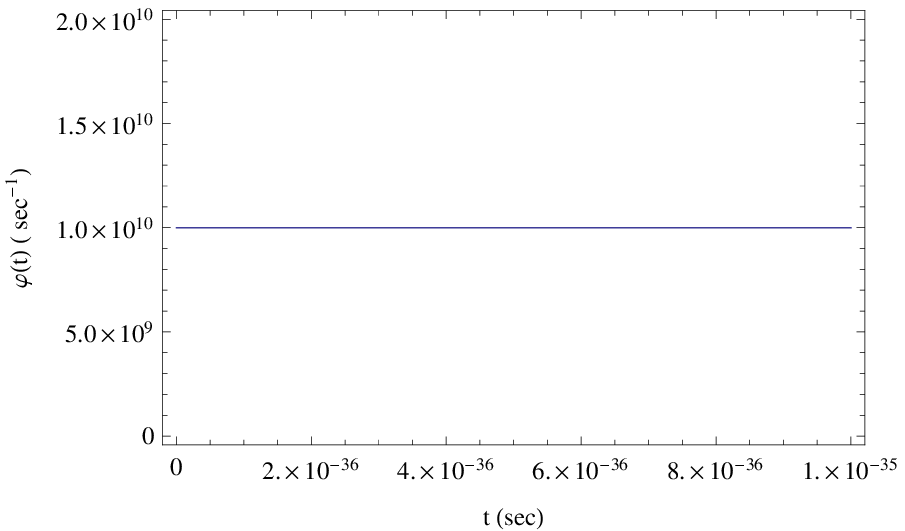}
\includegraphics[width=15pc]{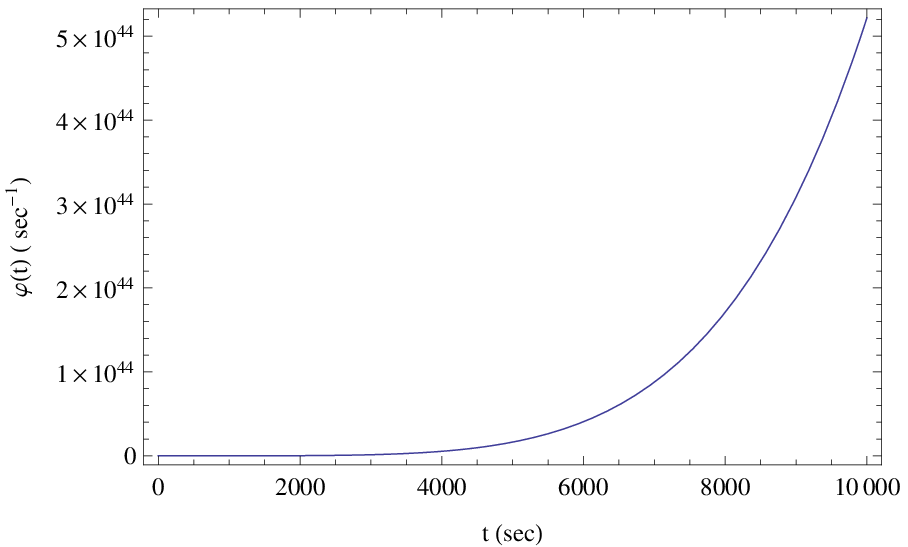}
\caption{\it{{The time dependence of the scalar field $\varphi (t)$,
with $c_1=10^{12}$, $c_2=10^{-20}$, $c_3=10^{10}$, $c_4=10^{-40}$,
$t_s=10^{-35}$sec,$\alpha=4/3$ and for the initial conditions
$\varphi (0)\simeq 10^{20}$ (sec $^{-1}$), $\varphi '(0)\simeq 10$ (sec$^{-2}$). }}}
\label{plot6}
\end{figure}

As a final task, we shall investigate the behavior of the potential
(\ref{scalarpotential}), for the chosen values of the parameters
(\ref{paramvalues}). Since we assumed that the scalar field
$\varphi$ does not evolve in a slow-roll way, it's dynamical
evolution is governed by the following non slow-roll equation of
motion,
\begin{equation}\label{nonslowrolleqnmotioncanonical}
\ddot\varphi + 3H\dot\varphi + \frac{\partial V(\varphi)}{\partial
\varphi}=0 \, ,
\end{equation}
with $V(\varphi)$ given in Eq. (\ref{scalarpotential}) and notice
that the difference between Eq.
(\ref{nonslowrolleqnmotioncanonical}) and Eq. (\ref{A9}), is traced
to the fact that the scalar field $\phi$ is non-canonical, while
$\varphi$ is canonical. We solve numerically the dynamical evolution
equation (\ref{nonslowrolleqnmotioncanonical}), and in Fig.
\ref{plot5}, we plotted the behavior of the potential as a function
of time, by choosing the initial conditions for the scalar field
$\varphi$ to be $\varphi (0)=10^{20}$ (sec$^{-1}$), and $\varphi' (0)=10^{-2}$ (sec$^{-2}$)
As it is obvious from Fig. \ref{plot5}, the scalar potential
dominates at early times and also remains quite large until $t\simeq
10^{-1}$sec, from which point it rapidly decays, and therefore the
second scalar field dominates the cosmological evolution. In order
to see how rapid is the decay of the potential $V(\phi (t))$, in
Table \ref{rapidflow}, we present some values of the scalar
potential $V(\phi (t))$, for various time points. Notice that at
$t=10^{-0.5}$sec, the potential decays quite rapidly and it is
practically zero at $t=0$.
\begin{table}[h]
\begin{center}
\begin{tabular}{|c|c|c|c|c|c|}
  \hline
  Time & $t\simeq 10^{-34}$ &  $t\simeq 10^{-1}$ & $t\simeq 10^{-0.49}$ & $t\simeq 10^{-0.48}$ & $t\simeq 10^{-0.47}$ \\
  \hline
  $V(\varphi (t))$ & $7.4\times 10^{59}$ & $8.5\times 10^{41}$ & $2\times 10^{24}$ & $516.3$ & $4.5\times 10^{-39}$\\
  \hline
\end{tabular}
\end{center}
\caption{\label{rapidflow}The scalar potential $V(\varphi (t))$ as a
function of cosmic time $t$, for the initial conditions $\varphi
(0)\simeq 10^{20}$ sec$^{-1}$, $\varphi '(0)\simeq 10^{-2}$ sec$^{-2}$, and also with
$c_1=10^{12}$, $c_2=10^{-20}$, $c_3=10^{10}$, $c_4=10^{-40}$,
$t_s=10^{-35}$sec,$\alpha=4/3$.}
\end{table}
An important remark is in order: The numerical analysis for the
scalar field $\varphi$ evolution, showed that the scalar field
$\varphi$ remains quite large for a wide range of time, therefore in
the previous analysis of the scalar field evolution $\chi(t)$, the
$\varphi$-dependent term can be safely neglected, since it is
negligible ($\varphi \rightarrow \infty$). The behavior of the
scalar field $\varphi (t)$ as a function of the cosmic time can be
seen in Fig. \ref{plot6}.
Finally, we have to note that the results are quite model dependent
and also strongly depend on the initial conditions and the values of
the parameters. However, the focus in this paper was mainly to
describe qualitatively the dynamical cosmological evolution and also
to indicate that the system is dynamically unstable. This is
the subject of the next section.

\section{Instability induced by Type IV singularity}

Having described the qualitative behavior of the model, we come to
the crucial part of our work, the instability of the system. As
we demonstrated in the previous section, the two scalar field model
with action (\ref{A1}) is described by the canonical scalar field
action of Eq. (\ref{approxma7}), before and during the inflationary
era, since the contribution of the second scalar field $\chi$ is
negligible during this era. However, the dynamical cosmological
evolution of the system is described by the Hubble rate
(\ref{IV1Bnoj}). In practice, the parameters are chosen in such a
way so that the term responsible for the singularity $\sim
(-t+t_s)^{\alpha}$, gives a very small contribution to the Hubble
rate, which is mainly driven by the first term in (\ref{IV1Bnoj}).
Since one of our main assumptions in this paper is that both scalar
fields do not respect the slow-roll evolution requirements, we shall
study the Hubble slow-roll parameters of Eq.
(\ref{hubbleslowrolloriginal}), for the Hubble rate
(\ref{IV1Bnoj}). A direct computation for the parameter $\epsilon_H$
yields,
\begin{equation}\label{epsilonhub}
\epsilon_H=\frac{\frac{c_1 c_3}{(c_2+c_3 t)^2}+c_4
(-t+t_s)^{-1+\alpha } \alpha }{\left(\frac{c_1}{c_2+c_3 t}+c_4
(-t+t_s)^{\alpha }\right)^2},{\,},
\end{equation}
while the parameter $\eta_H$ is equal to,
\begin{equation}\label{etahub}
\eta_H=-\frac{\frac{2 c_1 c_3^2}{(c_2+c_3 t)^3}+c_4
(-t+t_s)^{-2+\alpha } (-1+\alpha ) \alpha }{2
\left(\frac{c_1}{c_2+c_3 t}+c_4 (-t+t_s)^{\alpha }\right)
\left(-\frac{c_1 c_3}{(c_2+c_3 t)^2}-c_4 (-t+t_s)^{-1+\alpha }
\alpha \right)}{\,}.
\end{equation}
By looking Eq. (\ref{epsilonhub}), it is obvious that at early
times, the contribution of the terms $c_4 (-t+t_s)^{-1+\alpha }
\alpha$ and $c_4 (-t+t_s)^{\alpha }$ is negligible, compared to the
other terms. We can also verify this numerically, since for
$t=10^{-30}$sec, the term $c_4 (-t+t_s)^{-1+\alpha } \alpha$, is of
the order $\sim 10^{-51}$, while the other term in the numerator of
(\ref{epsilonhub}), namely $\frac{c_1 c_3}{(c_2+c_3 t)^2}$ is of the
order $10^{62}$.   Notice that we used the values for the parameters
defined in Eq. (\ref{paramvalues}). So practically we should neglect
the aforementioned two terms at early times, and consequently the
Hubble slow-roll parameter $\epsilon$ can be approximated by,
\begin{equation}\label{epsilonhub1}
\epsilon_H\simeq \frac{\frac{c_1 c_3}{(c_2+c_3
t)^2}}{\left(\frac{c_1}{c_2+c_3 t}\right)^2}=\frac{c_3}{c_1}{\,},
\end{equation}
and by taking into account the identifications we made in Eq.
(\ref{furtherconstriants}), we obtain that the early time
approximation of the Hubble slow-roll parameter $\epsilon_H$ for the
Hubble rate (\ref{IV1Bnoj}) is $\epsilon_H\simeq \frac{1}{q}$, which
is identical to the one we obtained in Eq.
(\ref{spedicslowrollind}). Hence, this result also verifies our
claim that the contribution of the slow-roll scalar $\chi$ at early
times is practically zero. Notice also that at $t=t_s$, the Hubble
slow-roll parameter is exactly $\epsilon=\frac{1}{q}$, without any
approximation, since both the terms $c_4 (-t+t_s)^{-1+\alpha }
\alpha$ and $c_4 (-t+t_s)^{\alpha }$, vanish, for the case of a Type
IV singularity, which occurs for $\alpha>1$. Hence, although the
Hubble slow-roll parameter $\epsilon$ is small and very well defined
for a Type IV singularity, the same issue of the graceful exit from
inflation persists, that is, inflation occurs, but never ends, the
slow-roll parameter is constant.

\subsubsection{Singular second Hubble slow-roll parameter and implications-Discussion on the results}

Let us now investigate what happens with the second Hubble slow-roll
parameter $\eta_H$, for the Hubble rate (\ref{IV1Bnoj}). As we can
see in Eq. (\ref{etahub}), in the case of a Type IV singularity at
$t=t_s$, and when $1<\alpha <2$, the numerator of $\eta_H$ becomes
divergent. This is very important, since this is an indication of
instability in
the exponential model. Moreover, the term responsible for the
singularity, that is, $c_4 (-t+t_s)^{-2+\alpha } (-1+\alpha )
\alpha$, is negligible for early cosmic times with $t>t_s$ and
$t<t_s$, and only becomes important at exactly the singularity point
$t=t_s$. The same applies for the corresponding term in the
denominator, so at early times and for $t>t_s$ and for $t<t_s$, the
second Hubble slow-roll parameter becomes, $\eta_H\simeq
\frac{1}{q}$, but at $t=t_s$, $\eta_H\rightarrow \infty$. Recall
that the second Hubble slow-roll parameter measures how long the
inflationary era is. 

So we have obtained a physically appealing
resulting picture, according to which, at early times, the two field scalar
model evolution is effectively identical to the one caused by a
single scalar field with potential (\ref{scalarpotential}), and with
slow-roll parameters that can be compatible with observations. In addition, the inflationary era ends violently at the point $t=t_s$
where a Type IV singularity occurs, at which point the second
slow-roll parameter $\eta_H$ becomes divergent. Since the point
$t_s$ can be arbitrarily chosen, this means that the inflationary evolution is unstable owing to this infinite singularity on the second Hubble
slow-roll parameters at every point we may wish.

The critical question is, what does this instability indicates? Could it indicate somehow that this could be a possible mechanism for a graceful exit from inflation in this exponential scalar model? The quantitative answer to these questions is not a trivial task however, so we discuss here qualitatively, the implications of the instability and the possibility to connect the instability to the graceful exit solution.

This instability mechanism we
propose for the ending of the inflationary era for the exponential
model (\ref{scalarpotential}), is quite different in spirit in
comparison to other existing mechanisms of ending of inflation, like
tachyonic instabilities \cite{encyclopedia} or via trace anomaly
\cite{sergeitraceanomaly}. Note that practically the singularity we
found, indicates some instability of the dynamical cosmological
system, but the qualitative description of this dynamical parameter
$\eta_H$ is that it measures when inflation ends, and the
singularity at $t=t_s$, definitely indicates strongly that the inflationary evolution is abruptly interrupted. So by having in mind that the only source of singularity is the second Hubble slow-roll index, it is rather tempting to ask if this instability could be actually indicate termination of the inflationary era. As we now demonstrate, this may be indeed occur.

In order to achieve this, we shall use the so-called slow-roll expansion, which is a generalization of the slow-roll approximation \cite{barrowslowroll}. In fact, the slow-roll expansion is a more accurate approach towards finding the inflationary solution, in comparison to the standard slow-roll approach. The slow-roll expansion was introduced in \cite{barrowslowroll}, and it is worth recalling in brief its basic features (for details see \cite{barrowslowroll}). Consider an inflationary solution that gives rise to a specific Hubble rate, $H(\varphi )$, where $\varphi$ is a canonical scalar field. Note that we shall for the moment adopt the notation of Ref. \cite{barrowslowroll}, and we shall express all the quantities to be used as functions of the canonical scalar, and in the end we express all the parameters as functions of the cosmic time, implicitly through the variables $\epsilon_H$ and $\eta_H$. In the context of the Hubble slow-roll expansion, an attractor inflationary solution is assumed, to which all the solutions that can be obtained from the potential slow-roll approximation, tend asymptotically. Note that the attractor inflationary solution is very important for the analysis that follows. Then, the single scalar field Friedmann equation reads,
\begin{equation}\label{frqwe}
H^2(\varphi )=\frac{8 \pi\kappa^2}{3}V(\varphi )\Big{(}1-\frac{1}{3}\epsilon_H (\varphi) \Big{)}^{-1}\, ,
\end{equation} 
where $\epsilon_H$ is the Hubble slow-roll parameter appearing in (\ref{hubbleslowrolloriginal}), but expressed as a function of the scalar field $\varphi$. By using the binomial theorem, it is possible to obtain the following perturbative expansion \cite{barrowslowroll} for the Friedmann equation (\ref{frqwe}),
\begin{align}\label{frqwe1}
& H^2(\varphi )\simeq \frac{8 \pi\kappa^2}{3}V(\varphi )\Big{(} 1+\epsilon_V-\frac{4}{3}\epsilon_V^2+\frac{2}{3}\epsilon_V\eta_V
\\ & \notag +\frac{32}{9}\epsilon_V^3+\frac{5}{9}\epsilon_V\eta_V^2-\frac{10}{3}\epsilon_V^2\eta_V+\frac{2}{9}\epsilon_V\xi^2_V+\mathcal{O}_4
 \Big{)}\, ,
\end{align} 
where we kept terms up to fourth order, and the parameters $\epsilon_V,\eta_V,\xi_V$ are defined in terms of the Hubble slow-roll parameters (\ref{hubbleslowrolloriginal}), as follows,
\begin{align}\label{shelloween}
& \eta_V=(3-\epsilon_H)^{-1}\Big{(}3\epsilon_H+3\eta_H-\eta_H^2-\xi_H^2\Big{)} ,
\\ & \notag \xi_V=(3-\epsilon_H)^{-1}\Big{(}27\epsilon_H\eta_H+9\xi_H^2-9\epsilon_H\eta_H^2-12\eta_H\xi_H^2-3\sigma_H^3+3\eta_H^2\xi_H^2+\eta_H\sigma_H^3 \Big{)}\, .
\end{align}
and also $\xi_H$ and $\sigma_H$ are given as functions of $\epsilon_H$ and $\eta_H$ below,
\begin{equation}\label{insertcode123sledgehammerenteraccept}
\xi_H^2=\epsilon_H\eta_H-\sqrt{\frac{1}{4\pi\kappa^2}}\sqrt{\epsilon_H}\eta_H',\,\,\,\sigma_H^3=\xi_H^2(2\epsilon_H-\eta_H)-\sqrt{\frac{1}{\kappa^2\pi}}\sqrt{\epsilon_H}\xi_H\xi_H'\, .
\end{equation}
Note that the prime denotes differentiation with respect to the scalar field $\varphi$ and also it is assumed that the solution $H(t)$ is the attractor of the theory. It is conceivable that the slow-roll approximation takes into account only the lowest order terms, while the expansion (\ref{frqwe1}) provides a better approximation to the inflationary attractor $H(t)$, which continues to describe the inflationary era, until the slow-roll expansion breaks down. As was also stressed by the authors of \cite{barrowslowroll}, the perturbative slow-roll expansion approximation breaks when the slow-roll parameters take large values, or if a singularity occurs. Using their example, for a potential of the form $\sim \varphi^2$, the expansion contains inverse powers of $\varphi$, and the inflationary era may end when the perturbative expansion is not valid anymore. This can occur definitely at the minimum of the potential at $\varphi=0$, where singularities appear in the expansion, but note that exit from inflation occurs earlier, owing to the fact that the expansion contains inverse powers of the scalar field $\varphi$.

The case we studied in this paper is perfectly described by what we just presented, that is, the exponential potential $\varphi$ drives the inflationary expansion until at some point a singularity occurs, so the solution is finally interrupted and the new attractor must be found. Practically this means that the inflationary attractor solution may be perfectly described by the perturbative slow-roll expansion solution, up to arbitrary order, until the perturbative expansion is violated, and the solutions exit from inflation. It is conceivable that the qualitative picture of our case goes as follows: The dynamical inflationary evolution is governed by the solution of the exponential scalar field, since the Type IV singular part of the Hubble rate (contribution of the second scalar), contributes very little to the dynamics. However, when the Type IV singularity time $t_s$ is reached, the second Hubble slow-roll parameter blows up and the expansion is not valid anymore, so this could indicate that inflation might end there. 

\begin{figure}[h]
\centering
\includegraphics[width=15pc]{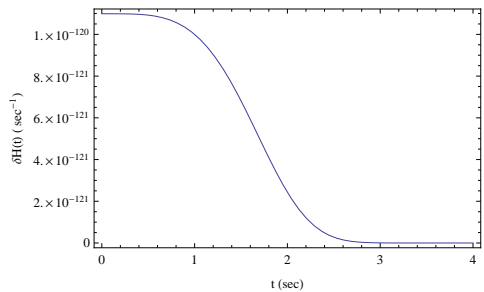}
\caption{\it{{The time dependence of the perturbation $\delta H(t)$,
with $c_1=10^{12}$, $c_2=10^{-20}$, $c_3=10^{10}$, $c_4=10^{-40}$,
$t_s=10^{-35}$sec,$\alpha=4/3$ and for the initial conditions
$\delta H(10^{-75})=10^{-120}$ ($sec^{-1}$) . }}}
\label{pap}
\end{figure}

Recall that the slow-roll expansion is constructed on the basic assumption that an attractor solution exists, before the exit from inflation. So as a final task, we shall prove that the solution $H(t)$, given in Eq. (\ref{IV1Bnoj}), can act as an attractor of the theory. Assuming that the attractor is,
\begin{equation}\label{attractor}
 H_0(t )= \frac{c_1}{c_2+c_3 t}+ c_4 \left( t_s - t \right)^\alpha
\end{equation} 
and that the dynamics is governed by the canonical scalar field, we linearly perturb the system $H=H_0+\delta H$, so at first order we get \cite{barrowslowroll},
\begin{equation}\label{forder}
-(\dot{H}_0)^2\dot{\delta H}=48 \pi \kappa^4 H_0\delta H\, ,
\end{equation}
with the solution to the differential equation being of the following form,
\begin{equation}\label{solform1}
\delta H=\delta H(t_i)\exp \Big{(}-12\kappa^2\int_{t_i}^{t}\frac{H_0(t)}{(\dot{H}_0(t))^2}\Big{)}\, .
\end{equation}
where $\delta H(t_i)$, the value of the perturbation at some early time $t_i$. By choosing the values of the parameter we used in the previous sections, and also by choosing $\delta H(t_i)$ to be very small, like for example $\delta H(t_i)=10^{-120}$ ($sec^{-1}$), with $t_i=10^{-75}$sec, we can solve numerically the integral in Eq. (\ref{solform1}). In Fig. \ref{pap}, we plot the behavior of the perturbation $\delta H$ as a function of time, where can see that the perturbation vanishes eventually, in an exponential way. Note however that this may depend on the initial conditions at the initial time $t_i$, but the exponential decay is the resulting picture again. Hence, the solution (\ref{attractor}), is the inflationary attractor of the theory, which is identical to the analytical expression that can result from the slow-roll expansion. However, at the Type IV singularity, the slow-roll expansion breaks down, and therefore this could indicate that the solution (\ref{attractor}) is no longer the inflationary attractor. Recall again that there is a clear distinction between the slow-roll approximation and the slow-roll expansion. In the present case, the slow-roll expansion breaks down at second order.

Before closing this section, we need to note that if the singularity
was Type III or even Type I, the Hubble slow-roll index $\epsilon_H$
would be singular at $t=t_s$, so practically we wouldn't be able to
even discuss about inflation, since the term $\sim
(t-t_s)^{\alpha-1}$, would make the Hubble slow-roll parameter
$\epsilon_H$, quite large. In order to have a taste of how large,
assume that $\alpha=-2$ (the Type I case according to our
classification) then for $t\sim 10^{-30}$sec, this term would be of
the order $10^{50}$, which is not negligible. Similar results hold
true for the case of the Type III, and for the Type II singularity,
but a full analysis is deferred to a future work, since this would
require to change all the numerical values we assumed for the
parameters and also we would have to change the initial conditions
for the scalar fields.

In the next section with shall study the stability of the
cosmological system of the two scalar fields when this is viewed as
a dynamical system.

\subsection{Cosmological dynamical system stability-Possible graceful exit from inflation via instabilities at the cost of phantom inflation}

In this section we shall study the stability of the cosmological
evolution caused by the two scalar fields, viewed as a dynamical
system. Particularly, we shall be interested in the stability of the
solution (\ref{A4}), which results from the reconstruction method we
used for two scalar fields. The cosmological evolution is given in
terms of the Hubble rate (\ref{IV1Bnoj}), and it is governed by the
FRW equations of Eq. (\ref{A2}). In order to better study the system
of cosmological equations as a dynamical system, we shall introduce
the quantities $X_{\phi}$, $X_{\chi}$ and $\bar{Y}$, which are
related to the scalar fields $\phi$ and $\chi$ as follows,
\begin{equation}\label{newvariables}
X_{\phi}=\dot{\phi}\, , \quad  X_{\chi}=\dot{\chi}\, , \quad
\bar{Y}=\frac{\tilde{f}(\phi ,\chi)}{H}\, ,
\end{equation}
and in view of Eq. (\ref{newvariables}), the FRW equations
(\ref{A2}) can be written as,
\begin{align}\label{neweqs}
& \frac{\mathrm{d}X_{\phi}}{\mathrm{d}N}=\frac{\omega '(\phi)\left
(X_{\phi}^2-1\right )}{2\omega(\phi)H}-3\left
(X_{\phi}-\bar{Y}\right )\nn &
\frac{\mathrm{d}X_{\chi}}{\mathrm{d}N}=\frac{\eta '(\chi)\left
(X_{\chi}^2-1\right )}{2\eta(\chi)H}-3\left (X_{\chi}-\bar{Y}\right
) \nn &\frac{\mathrm{d}\bar{Y}}{\mathrm{d}N}=\frac{3 X_{\phi}
X_{\chi}\left (1-\bar{Y}^2\right
)}{X_{\phi}+X_{\chi}}+\frac{\dot{H}}{H^2}\frac{X_{\phi}X_{\chi}+1-\bar{Y}(X_{\phi}+X_{\chi})}{X_{\phi}+X_{\chi}}
\, .
\end{align}
Notice that the solution of Eq.~(\ref{A4}), can be expressed in
terms of the variables (\ref{newvariables}), as follows,
\begin{equation}\label{valuesofvariables}
X_{\phi}=1\, , \quad X_{\chi}=1\, , \quad \bar{Y}=1\, .
\end{equation}
By performing the following linear perturbations of the dynamical
variables (\ref{newvariables}),
\begin{equation}\label{dynamicalpertyrbationsii}
X_{\phi}=1+\delta X_{\phi} \, , \quad X_{\chi}=1+\delta X_{\chi}\, ,
\quad \bar{Y}=1+\delta \bar{Y}\, ,
\end{equation}
we obtain the following dynamical system which describes the
evolution of the perturbations, namely,
\begin{equation}\label{dynamicalsystemiii}
\frac{\mathrm{d}}{\mathrm{d}N}\left(
\begin{array}{c}
  \delta X_{\phi} \\
  \delta X_{\chi} \\
  \delta \bar{Y}\\
\end{array}
\right)=\left(
\begin{array}{ccc}
 -\frac{\omega' (\phi)}{H\omega (\phi)}-3 & 0 & 3
 \\ 0 &-\frac{\eta' (\chi)}{H\eta (\chi)}-3 & 3 \\
 0 & 0 & -3-\frac{\dot{H}}{H^2} \\
\end{array} \right)\left(
\begin{array}{c}
   \delta X_{\phi} \\
  \delta X_{\chi} \\
  \delta \bar{Y}\\
\end{array}
\right) \, .
\end{equation}
The eigenvalues corresponding to the matrix appearing in the
dynamical system above, are equal to,
\begin{equation}\label{eigenvaluesdoblescal}
M_{\phi}=-\frac{\omega' (\phi)}{H\omega (\phi)}-3\, , \quad
{\,}M_{\chi}=-\frac{\eta' (\chi)}{H\eta (\chi)}-3\, , \quad
M_{\bar{Y}}=-3-\frac{\dot{H}}{H^2} \, .
\end{equation}
Hence, by substituting the Hubble rate (\ref{IV1Bnoj}), we obtain,
\begin{align}\label{eigendoublehill}
& M_{\phi}=-3-\frac{2 c_3}{\left(\frac{c_1}{c_2+c_3 t}+c_4
(-t+t_s)^{\alpha }\right) (c_2+c_3 \phi )} \, ,\nn &
M_{\chi}=-3-\frac{-1+\alpha }{\left(\frac{c_1}{c_2+c_3 t}+c_4
(-t+t_s)^{\alpha }\right) (t_s-\chi )} \, ,\nn &
M_{\bar{Y}}=-3+\frac{\frac{c_1 c_3}{(c_2+c_3 t)^2}+c_4
(-t+t_s)^{-1+\alpha } \alpha }{\left(\frac{c_1}{c_2+c_3 t}+c_4
(-t+t_s)^{\alpha }\right)^2} \, ,
\end{align}
From the theory of dynamical systems stability \cite{jost} it is
known that when all the eigenvalues are negative, the dynamical
system is stable, so the evolution of perturbations
(\ref{dynamicalpertyrbationsii}) is stable. When one eigenvalue is
positive and the other two negative, then the dynamical system is
unstable, when two are positive and one negative, the system is
stable, and when three eigenvalues are positive, then the dynamical
system is unstable. Finally when one or more eigenvalues are zero,
then we have a saddle point in the dynamical evolution of
perturbations. As it is clear from Eq. (\ref{eigendoublehill}), for
the choice of the parameters as in Eq. (\ref{paramvalues}) both
$M_{\phi}$ and $M_{\bar{Y}}$ are negative, but for the eigenvalue
$M_{\chi}$, it is always positive, if the value of $\chi(t)$ is
larger from $t_s$. This however strongly depends on the initial
conditions for the field $\chi (t)$. For the initial conditions we
chose (\ref{initialcondscalarchi}) in the previous sections, this is
always true since $\chi(0)=10^{-25}$, and the scalar field $\chi
(t)$ increases with time, as we evinced in a previous section.
Therefore, in our case the dynamical system always develops an
instability, with this feature probably indicating that the
inflationary solution driven by the scalar field $\varphi $ at early
times is not the final attractor of the system, at least for the
initial conditions we chose. However, the full study of the
cosmological dynamical system exceeds the purposes of this paper,
since our aim is to qualitatively stress the fact that, the Type IV
singularity in some way makes the system unstable and the
inflationary solution of the system at early times is not the final
attractor of the system.

Finally, we should note that if the initial conditions of the scalar
field $\chi(t)$, were chosen to be smaller than the value of $t_s$,
or if in our case $t_s$ is chosen to be larger than $10^{-25}$, so
that $\chi(0)<t_s$, then it is possible that, as $\chi$ increases,
at some point $t_1$ corresponding to the crossing point
$t_s=\chi(t_1)$, the eigenvalue $M_{\chi}$ becomes infinite. This
infinity may also indicate an instability of the dynamical system,
thus indicating an abrupt change of the final attractor solution of
the scalar field $\varphi$. From a mathematical point of view, this
dynamical system analysis is a difficult issue and should be
scrutinized in order to come to rigid conclusions. From this brief
study we keep only the result that the dynamical system of
perturbations of the cosmological parameters $X_{\phi}$, $X_{\chi}$
and $\bar{Y}$, is unstable.

\section{Late-time evolution of the two scalar field model}

In the previous sections we thoroughly studied the early time
behavior of the two scalar field model, but we did not address the
late-time evolution study of the model. Here we study the late-time
behavior of the two scalar model, in terms of the effective equation
of state (EoS) of the model. Before getting into the detailed
analysis, let us recall the essential features of the two scalar
field model we used, assuming that the parameters are as these
appear in Eq. (\ref{paramvalues}). Firstly, we assumed that the Type
IV singularity occurs at early times, specifically at $t=t_s$, and
secondly, the two scalar fields are required not to obey the
slow-roll conditions. As we evinced, for the choice of the variables
as in Eq. (\ref{paramvalues}) and also by assuming that the initial
conditions of the scalar field $\chi$ are taken as in Eq.
(\ref{initialcondscalarchi}), we demonstrated that the scalar field
$\chi$ has a negligible contribution to the cosmological evolution
of the model at early times, but dominates the cosmological
evolution at late times. As we now show, this behavior can also be
seen by studying the EoS of the two scalars system. The EoS
parameter $w_{eff}$ for the two scalar model is given by,
\begin{equation}
\label{eos} w_\mathrm{eff}=\frac{p}{\rho}=-1-\frac{2\dot{H}}{3H^2}
\, .
\end{equation}
By using the form of the Hubble rate given in Eq. (\ref{IV1Bnoj}),
the EoS reads,
\begin{equation}\label{weeffnewstudy}
w_{\mathrm{eff}}=-1-\frac{2 \left(-\frac{c_1 c_3}{(c_2+c_3 t)^2}-c_4
(-t+t_s)^{-1+\alpha } \alpha \right)}{3 \left(\frac{c_1}{c_2+c_3
t}+c_4 (-t+t_s)^{\alpha }\right)^2}\, .
\end{equation}
The above relation (\ref{weeffnewstudy}) will be the starting point
of our analysis. We shall investigate the behavior of the EoS at
early and at late times. We start off with early time, in which
case, the contribution of the terms $\sim c_4 (-t+t_s)^{-1+\alpha }
\alpha$ and $c_4 (-t+t_s)^{\alpha }$ appearing in the numerator and
the denominator of the fraction in Eq. (\ref{weeffnewstudy}), is
negligible compared to the other terms. Also, when $t\simeq t_s$,
these are identically zero, for the value of the parameter $\alpha $
we chose. Therefore, by neglecting these, the EoS at early times
reads,
\begin{equation}\label{weffneartypeiv}
w_{\mathrm{eff}}\simeq -1+\frac{2 c_3}{3 c_1}\, ,
\end{equation}
which clearly describes quintessential acceleration, different from
de Sitter though, since in our case $\frac{c_3}{c_1}=\frac{1}{q}$
(see Eq. (\ref{furtherconstriants})) and we took $q=100$, so it
cannot be considered nearly de Sitter, but purely quintessential
acceleration. Hence the inflationary era is a quintessence era,
without the possibility of evolving into a phantom evolution.  With
regards to the late-time era, it is conceivable that as $t$ grows
larger, approaching the present time era $t_p\simeq 4.25 \times
10^{17}$sec, the terms $\sim c_4 (-t+t_s)^{-1+\alpha } \alpha$ and
$c_4 (-t+t_s)^{\alpha }$ will start to dominate the evolution,
therefore the EoS at late-time reads,
\begin{equation}\label{lateeos}
w_{\mathrm{eff}}\simeq -1+\frac{2 (-t+t_s)^{-1-\alpha } \alpha }{3
c_4}\, ,
\end{equation}
and since $\alpha =4/3$ and also $t_s\ll t$, then,
\begin{equation}\label{largetaprrox}
(-t+t_s)^{\alpha-1}\simeq (-t)^{\alpha-1}=-t^{\alpha-1}\, .
\end{equation}
Consequently, the EoS (\ref{lateeos}) becomes at late times,
\begin{equation}\label{newfinaleos}
w_{\mathrm{eff}}\simeq -1-\frac{2 t^{-1-\alpha } \alpha }{3 c_4}\, .
\end{equation}
The EoS (\ref{newfinaleos}) describes nearly phantom acceleration,
with nearly meaning that the acceleration is slightly phantom.
Indeed, since we chose $c_4=10^{-40}$, and also owing to the fact
that we consider late times, when $t$ is of the order of the present
time, that is $t\simeq 4\times 10^{17}$sec, then the EoS is
approximately equal to $w_{\mathrm{eff}}\simeq -1.06545$. Note that
observations indicate that the EoS at present time or in near future
may crossed the phantom divide \cite{phantom}, so this result is
quite interesting.

In conclusion, in the context of the two scalar field model that
realizes a Type IV singularity at early times, we were able to
describe early and late-time acceleration in a unified way, by using
the same theoretical framework. One interesting scenario that we did
not address here is to study what is the impact on the two scalar
field model we used, if the Type IV singularity occurs at late-time.
However, this would require to change completely the constraints we
used, so we hope to address this in the future.

\section{Type IV singularity as a source of instability for other scalar models-Universal description for non-slow-roll evolution}

From all the previous sections it is conceivable that the instability in scalar models caused by a Type IV singularity, can
be achieved if the slow-roll conditions for the scalar fields are not
taken into account. Then, it is possible that the Hubble slow-roll
parameters are small during the inflationary era, but these blow-up
at the Type IV singularity, where inflation is abruptly interrupted. So by saying that
the slow-roll condition is not taken into account, it is meant that
the indicators for the inflationary dynamics are not the potential
slow-roll parameters \cite{barrowslowroll}, but the Hubble slow-roll
parameters \cite{barrowslowroll}. In this section we briefly
investigate if this instability mechanism we proposed in this
paper, can also generate have implications on other scalar models if
these are put into proper context. We shall study two quite popular
scalar models, a large field inflation model
\cite{encyclopedia,noo1} and a modified $R^2$ inflation model
\cite{encyclopedia,noo3}. We start off with the large field
inflation model, which we studied in detail in \cite{noo1} and we
were able to associate it to a Type IV singular evolution, by using
a single scalar field and not two. Let us recall the essential
information for the Type IV singular evolution for the large field
inflation and for details the reader is referred to \cite{noo1}. The
Hubble rate that describes the Type IV singular evolution is equal
to,
\begin{equation}\label{hubnew1}
H(t)=f_0\left(t_s-t\right)^{\alpha}\, ,
\end{equation}
so for $\alpha >1$ a Type IV singularity occurs.  Consider the
single scalar field action, \be \label{ma7newsection} S=\int d^4 x
\sqrt{-g}\left\{ \frac{1}{2\kappa^2}R -
\frac{1}{2}\omega(\phi)\partial_\mu \phi
\partial^\mu\phi - V(\phi) + L_\mathrm{matter} \right\}\, .
\ee
 where the function $\omega(\phi)$ stands for the kinetic function and
$V(\phi)$ is the scalar potential of the scalar field $\phi$. Using
Eq.~(\ref{ma7newsection}), the effective energy density and the
effective pressure of the non-canonical scalar field $\phi$ can be
written,
 \be \label{ma8newsection} \rho = \frac{1}{2}\omega(\phi){\dot \phi}^2
+ V(\phi)\, ,\quad p = \frac{1}{2}\omega(\phi){\dot \phi}^2 -
V(\phi)\, . \ee
 Then in view of Eq. (\ref{ma8newsection}) the scalar
potential $V(\phi)$ and the kinetic function $\omega(\phi)$ can be
expressed in terms of the Hubble rate as follows,
 \be \label{ma9}
\omega(\phi) {\dot \phi}^2 = - \frac{2}{\kappa^2}\dot H\, ,\quad
V(\phi)=\frac{1}{\kappa^2}\left(3H^2 + \dot H\right)\, . \ee
 The
single scalar field reconstruction method is based on the crucial
assumption that the kinetic function $\omega(\phi)$ and the scalar
potential $V(\phi)$, can be expressed as functions of a single
function $f(\phi)$, as follows,
 \be \label{ma10} \omega(\phi)=-
\frac{2}{\kappa^2}f'(\phi)\, ,\quad
V(\phi)=\frac{1}{\kappa^2}\left(3f(\phi)^2 + f'(\phi)\right)\, . \ee
For further details on this method, see \cite{Nojiri:2005sx,noo1}.
Consequently, the FRW equations for the single scalar read
\cite{Nojiri:2005sx,noo1},
 \be \label{ma11} \phi=t\, ,\quad H=f(t)\,
. \ee

The kinetic function and scalar potential for the scalar $\phi$, in
the case that the Hubble rate is given as in (\ref{hubnew1}), near
$t=t_s$, are equal to,

\be \label{IV4POTNEW} \omega(\phi) = \frac{2\alpha f_0}{\kappa^2}
\left( t_s - \phi \right)^{\alpha - 1} \, , \quad V(\phi) \sim -
\frac{\alpha f_0 }{\kappa^2} \left( t_s - \phi \right)^{\alpha - 1}
\, , \ee By canonically transforming the single scalar field $\phi$
(see also appendix), we can have the potential in the following
form,

\be \label{IV6potnew} V(\varphi) \sim - \frac{\alpha f_0 }{\kappa^2}
\left\{ - \frac{\kappa \left(\alpha+ 1 \right) }{2 \sqrt{2\alpha
f_0}} \varphi \right\}^{\frac{2 \left( \alpha - 1 \right)}{\alpha +
1}}\, , \ee
where the field $\varphi$ is canonical. The potential
(\ref{IV6potnew}), is known to describe large field inflation
\cite{Barrow:2015ora,encyclopedia}. The observational implications
of the potential (\ref{IV6potnew}), are quite appealing as we
evinced in \cite{noo1}, in the case that the singularity occurs at
the end or after inflation, but in the context of a slow-roll
evolution. Now consider that we abandon the slow-roll evolution
requirement and in addition assume that the singularity can occur at an
arbitrary time instance. The Hubble slow-roll parameters for the
Hubble rate (\ref{hubnew1}), are equal to,
\begin{equation}\label{epsilonhublargefield}
\epsilon_H=\frac{(-t+t_s)^{-1-\alpha } \alpha
}{f_0},{\,}{\,}{\,} \eta_H=\frac{(-t+t_s)^{-1-\alpha }
(-1+\alpha )}{2 f_0}{\,}.
\end{equation}
Therefore by choosing $f_0$ to be enormously large, larger than
$(-t+t_s)^{-1-\alpha }$, that is, $f_0 \gg
(-t+t_s)^{-1-\alpha }$, for $t\simeq t_s$, then both the
Hubble slow-roll parameters $\epsilon_H$ and $\eta_H$ can be
significantly smaller than unity, $\epsilon_H,\eta_H \ll 1$, in the
neighborhood of $t\sim t_s$. But for a Type IV singularity
($\alpha>1$), both the Hubble slow-roll parameters blow up to
infinity at $t=t_s$, where the singularity occurs. Therefore, even
in the non-slow-roll approximation, this could be an indication that
inflation is interrupted severely due to the existence of an singular
instability. Hence, our mechanism can work for other potentials too,
apart from the exponential we explored in the previous sections.
Notice that $f_0$ is a free parameter, so it can be chosen
appropriately in order concordance with Planck data can be achieved,
at least in the neighborhood of the Type IV singularity where inflation can be chosen to end.

Before closing, we demonstrate in brief that the same mechanism can
create grateful exit from inflation for nearly $R^2$ potentials
which can generate a singular evolution, which we studied in
\cite{noo3}. As we evinced in \cite{noo3}, the Hubble rate for the
Type singularity generating nearly $R^2$ potentials reads,

\be \label{IV1Bnojgrexit} H(t) =  \frac{c_1}{c_2+c_3 t}+c_4+ c_5
\left( t_s - t \right)^\alpha\, , \ee
and consequently, the first Hubble slow-roll parameter $\epsilon_H$
is equal to,
\begin{equation}\label{epsilonhubnew}
\epsilon_H=-\frac{-\frac{c_1 c_3}{(c_2+c_3
t)^2}-c_5 (-t+t_s)^{-1+\alpha } \alpha
}{\left(c_4+\frac{c_1}{c_2+c_3 t}+c_5
(-t+t_s)^{\alpha }\right)^2}{\,},
\end{equation}
while the parameter $\eta_H$ is equal to,
\begin{equation}\label{etahubnew}
\eta_H=-\frac{\frac{2 c_1 c_3^2}{(c_2+c_3
t)^3}+c_5 (-t+t_s)^{-2+\alpha } (-1+\alpha ) \alpha }{2
\left(c_4+\frac{c_1}{c_2+c_3 t}+c_5
(-t+t_s)^{\alpha }\right) \left(-\frac{c_1
c_3}{(c_2+c_3 t)^2}-c_5
(-t+t_s)^{-1+\alpha } \alpha \right)}{\,}.
\end{equation}
Hence, for $1<\alpha<2$, the second slow-roll parameter blows up at
the Type IV singularity. Therefore, in this case too, the Type IV
singularity induced instability can work in the
same way as in the exponential model we studied in the previous
sections. Note that the model studied in \cite{noo3} and the one we
studied in this paper have many qualitative similarities, except
that in \cite{noo3} we used the slow-roll approximation for the
canonical scalar $\varphi$.

\section{Concluding remarks}

In this paper we included a Type IV singularity in the dynamical
evolution of a quite well studied exponential scalar model, which is
known to be compatible with the Planck data, when the spectral index
of primordial perturbations is considered. In order to successfully
incorporate the Type IV singularity we used two scalar fields, and
by employing very well known scalar reconstruction techniques, we
managed to produce the Type IV singular behavior appearing in the
Hubble rate of the model. The scalar exponential model we studied
however, is known to suffer from the graceful exit from inflation
problem, but in the context of our model, we provided an indication on how inflation might end in this model. However, this issue has to be thoroughly addressed, a task which is beyond the scopes of this introductory paper on these issues.

In order
to generate an instability, we assumed that both the scalar fields, one being
canonical and one non-canonical, do not satisfy the slow-roll
requirements. Also, the values of the parameters and the initial
conditions on the fields are chosen in such a way, so that the model
exhibits the following features:
\begin{itemize}

\item The Type IV is assumed to occur at early times. In the context of our model, this will indicate the end of inflation in the model.

\item At early times, the canonical scalar, which is the one that corresponds to the exponential scalar model,
dominates the cosmological evolution, with the scalar potential and
kinetic term of the non-canonical field being neglected. Special
emphasis was given on the inflationary era.

\item The non-canonical scalar never becomes a phantom scalar, so singular phantom evolution is avoided.

\end{itemize}
As a side effect of our model, the late-time evolution is governed
by the non-canonical scalar, as we analyzed in detail. In fact, in
the context of our model, late-time acceleration and early-time
acceleration can be described in a unified way, as we evinced.

For our two scalar field model, since the slow-roll condition is not
respected by both scalar fields, the indicators that determine the
inflationary process, if it occurs and when it ends, are the Hubble
slow-roll parameters. By calculating these for a Type IV
singularity, we obtained the quite appealing result that at the Type
IV singularity, the second Hubble slow-roll parameter, which is
known to indicate the end of inflation, diverges at the Type IV
singularity. As we claimed, this might be an indication that
inflation ends at the point of the Type IV singularity. Away from
the singularity however, and always referring to early times, the
Hubble slow-roll indices are approximately (and in some cases
exactly) equal to the ones corresponding to the exponential scalar
model solely. This in some way also validates our claims about the
evolution of the two scalar field system. In addition, we supported
numerically
 and verified in a semi-quantitative way our theoretical
claims. We need to note though that in order to obtain some
quantitative results, the analysis must be much more rigid and also
someone should examine the dynamical system of the cosmological
equations exactly. Our aim was to indicate in a qualitative and
semi-quantitative way, the fact that a Type IV singularity might be
responsible for the graceful exit from inflation, to certain
cosmological models. Finally, we studied analytically the stability
of the cosmological equations, when these are viewed as a dynamical
system, by examining the dynamical system of linear perturbations of
certain parameters, related to some physical quantities. As we
demonstrated, the system is unstable, with this instability possibly
indicating the fact that the initial solution corresponding to the
canonical scalar field is not an attractor of the dynamical system.
Indeed, this was the case, since the solution lasts up to the point
that the Type IV singularity occurs, with the duration of the
solution being quantified by the second Hubble slow-roll parameter.

We need to note that the effects of the Type IV singularity on the
cosmological evolution of the model are somehow unique since only
the second slow-roll parameter diverges and not the first. One
should however examine the rest of finite time singularities, with
the most appealing, from a phenomenological point of view, being the
Type II. Since finite time singularities frequently occur in
cosmology, the phenomenological implications of these have to be
understood in detail, especially the effects of the mildest ones,
like the Type IV and Type II. These could be the link between a
quantum theory of cosmology and the classical cosmological theory,
with the most appealing candidate of a quantum cosmology theory
being Loop Quantum Cosmology \cite{LQC}. Finally, it is worth to study the evolution of the model we studied in this paper, in the context of theories with generalized equation of state. For a recent study see \cite{fin}.

\section*{Acknowledgments}

The work of S.O. is supported by MINECO (Spain), projects
FIS2010-15640 and FIS2013-44881.

\section*{Appendix}

Here we discuss in brief why the incorporation of the Type IV
singularity in the cosmological evolution of a single scalar field
is very difficult, not to say impossible. Consider the following
generic non-canonical scalar field action, \be \label{ma7} S=\int
d^4 x \sqrt{-g}\left\{ \frac{1}{2\kappa^2}R -
\frac{1}{2}\omega(\phi)\partial_\mu \phi
\partial^\mu\phi - V(\phi)  \right\}\, .
\ee with the function $\omega(\phi)$ being the kinetic function and
the scalar potential is $V(\phi)$. For the flat FRW background of
Eq. (\ref{JGRG14}), the energy density and the corresponding
pressure are equal to, \be \label{ma8} \rho =
\frac{1}{2}\omega(\phi){\dot \phi}^2 + V(\phi)\, ,\quad p =
\frac{1}{2}\omega(\phi){\dot \phi}^2 - V(\phi)\, . \ee Therefore,
the potential $V(\phi)$ and the kinetic function $\omega(\phi)$ can
be expressed in terms of the Hubble rate in the following way, \be
\label{ma9} \omega(\phi) {\dot \phi}^2 = - \frac{2}{\kappa^2}\dot
H\, ,\quad V(\phi)=\frac{1}{\kappa^2}\left(3H^2 + \dot H\right)\, .
\ee By making the transformation (\ref{ma13}), we can rewrite the
action of Eq. (\ref{ma7}), in terms of the canonical scalar field
$\varphi$, since the kinetic term becomes, \be \label{ma13b}
 - \omega(\phi) \partial_\mu \phi \partial^\mu\phi
= - \partial_\mu \varphi \partial^\mu\varphi\, . \ee and finally the
action of Eq. (\ref{ma7}) reads, \be \label{ma7einfram12e} S=\int
d^4 x \sqrt{-g}\left\{ \frac{1}{2\kappa^2}R -
\frac{1}{2}\partial_\mu \varphi
\partial^\mu\varphi - V(\varphi) \right\}\, .
\ee The technique that simplifies to a great extent the
incorporation of finite time singularities, is the
scalar-reconstruction technique developed in Refs.
\cite{Nojiri:2005pu,Capozziello:2005tf}, which is based on the
assumption that the kinetic function and the scalar potential can be
written in terms of a function $f(\phi)$, in the following way, \be
\label{ma10} \omega(\phi)=- \frac{2}{\kappa^2}f'(\phi)\, ,\quad
V(\phi)=\frac{1}{\kappa^2}\left(3f(\phi)^2 + f'(\phi)\right)\, , \ee
Therefore, if we assume that no other matter fluids are present in
the cosmological equations, the field equations (\ref{ma8}) take the
following form, \be \label{ma11} \phi=t\, ,\quad H=f(t)\, . \ee As
in the $R^2$ inflation case, which we studied in \cite{noo3}, our
inability to incorporate the finite time singularity for the
potential (\ref{scalarpotential}), is traced on the fact that the
potential has an exponential functional dependence with respect to
the scalar field $\varphi$. In the case of a Type IV singularity, a
quite general form of the function $f(t)$ that can describe the Type
IV singular evolution, has the form,
\begin{equation}\label{f}
f(t)=c(t-t_s)^b+f_1(t)
\end{equation}
with $f_1(t)$ being the non-canonical scalar-tensor reconstruction
function that produces the scalar potential (\ref{scalarpotential}).
Equivalently, $f(t)$ can be equal to,
\begin{equation}\label{newf}
f(t)=f_1(t)(t-t_s)^b ,{\,}{\,}{\,}f(t)=e^{(t-t_s)^{\alpha}}+c
\end{equation}
with $c$ being an arbitrary constant. For both the cases (\ref{f})
and (\ref{newf}), it is very difficult to incorporate the Type IV
singularity for the following reasons:

\begin{itemize}

\item The function $f(t)$ must be chosen in such a way, so that it is possible to solve the
integral $\varphi =\int^{\phi} \sqrt{f'(\phi )} \mathrm{d}\phi$, in
terms of $\phi =\phi (\varphi)$. This is not so easy to be done for
the functions $f(t)$ of  Eqs. (\ref{f}) and (\ref{newf}).

\item In the case that an ansatz is found and the function  $f(\phi)$ is found, so that $\phi =\phi (\varphi)$
can be explicitly solved, then even in this case, the resulting
scalar potential becomes severely constrained and the potential
(\ref{scalarpotential}) cannot be easily reproduced even in parts.
In order to see this, suppose that $\phi(\varphi)$ is found in
explicit form, then the corresponding scalar potential $V(\phi
(\varphi))$ would be equal to, \be \label{ma100101} V(\phi
(\varphi))=\frac{1}{\kappa^2}\left(3f(\phi(\varphi))^2 + f'(\phi
(\varphi))\right)\, , \ee and this potential must be of the
following form,
\begin{equation}\label{defoeme1}
V(\phi (\varphi))=V_0e^{-\sqrt{\frac{2}{q}}\kappa \varphi}+V_1(\phi
(\varphi))
\end{equation}
which is a rather formidable task to do, since an ansatz is needed.
Work is in progress though to find such an ansatz.

\end{itemize}

\end{document}